\documentclass{llncs}
\usepackage{latexsym,color,graphics,times}
\usepackage{diagrams}
\usepackage{url}
\usepackage{epsfig}
\usepackage{amssymb}
\usepackage{amsmath}
\usepackage{amsfonts}

\begingroup
\catcode`\~=11
\gdef\urltilde{\lower 0.6ex\hbox{~}}
\endgroup

\newcommand{\A}{\mathcal{A}} 
\newcommand{\C}{\mathcal{C}} \newcommand{\D}{\mathcal{D}}

\newcommand{\K}{\mathcal{K}} \renewcommand{\L}{\mathcal{L}}
\newcommand{\M}{\mathcal{M}} 
 \renewcommand{\P}{\mathcal{P}}
 
 \newcommand{\T}{\mathcal{T}}

\title{Data Base Mappings and Monads: (Co)Induction}
\date{}

\author{Zoran Majki\'c}
\institute{ International Society for Research in Science and Technology\\
 PO Box 2464 Tallahassee, FL 32316 - 2464 USA\\
 \email{majk.1234@yahoo.com},\\
~~~~http://zoranmajkic.webs.com/}

\newtheorem{theo}{Theorem}
\newtheorem{propo}{Proposition}
\newtheorem{coro}{Corollary}
\begin{document}


\maketitle
\begin{abstract}
In this paper we presented the semantics of database mappings in the
relational $DB$ category based on the power-view monad $T$ and
monadic algebras. The objects in this category are the
database-instances (a database-instance is a set of n-ary relations,
i.e., a set of relational tables as in standard RDBs). The morphisms
in $DB$ category are used in order to express the semantics of
view-based Global and Local as View (GLAV) mappings between
relational databases, for example those used in Data Integration
Systems. Such morphisms in this $DB$ category are not functions but
have the complex tree structures based on a set of complex query
computations between two database-instances.  Thus $DB$ category, as
a base category for the semantics of databases and mappings between
them, is different from the  $Set$ category used dominantly for such
issues, and needs the full investigation of its properties.\\
  In this paper we presented
another contributions for an intensive exploration of properties and
semantics of this category, based on the power-view monad $T$ and
the Kleisli category for databases. Here we stressed some Universal
algebra considerations based on monads and relationships between
this $DB$ category and the standard $Set$ category. Finally, we
investigated the general algebraic and induction properties for
databases in this category, and we defined the \emph{initial}
monadic algebras for database instances.
    \end{abstract}
%
\section{Introduction}
The computational significance of monads has been stressed
\cite{Mogg89,Mogg91} in suggestions that they may help in
understanding programs "as functions from values to computations".
The idea, roughly, is to give a denotational semantics to
computations, and it suggests an alternative to the conceptual gap
between the intensional (operational) and the extensional
(denotational) approach to the semantics of programming languages.\\
The idea of a monad, based on an endofunctor $T$ for a given
category, as a model for computations is that, for each set of
values of type $A$, $TA$ is the object of computations of "type
$A$".\\
Let us explain in which way we can use such a denotational
semantics, based on monads, in the case of relational databases.
  It is well known that the relational databases are  complex structures, defined by sets of n-ary relations, and the
mappings between them are based on sets of view-mappings between the
source database $A$ to the target database $B$.  We consider the
views as an universal property for databases
  (possible observations of the information contained in some
  database).\\ We assume \emph{a
view} of a database $A$ the  relation (set of tuples) obtained by a
"Select-Project-Join + Union" (SPJRU) query $q(\textbf{x})$ where
$\textbf{x}$ is a list of attributes of this view. We denote by
$\L_A$ the set of all such queries over a database A, and by
$\L_A/_\approx$ the quotient term  algebra obtained by introducing
the equivalence relation $\approx$, such that $q(\textbf{x}) \approx
q'(\textbf{x})$ if
  both queries result with the same relation (view). Thus,
a view can be equivalently considered as a  \emph{term} of this
quotient-term algebra $\L_A/_\approx$ with carrier set of relations
in $A$ and a finite arity of their operators, whose computation
returns with a set of tuples of this view. If this query is a finite
term of this algebra it is
    called a "finitary view". Notice that a finitary view can have an infinite number of tuples
    also.\\
 Such an  \emph{instance level} database category $DB$ has been introduced first time in Technical report ~\cite{Majk03}, and used also in
 \cite{Majk03f}. General information about categories the reader can find in classic books ~\cite{McLn71}, while more information about this
 particular database category $DB$,
  with set of its objects $Ob_{DB}$ and set of its morphisms
  $Mor_{DB}$, are recently presented  in \cite{Majk04AOT}.
  In this paper we will only emphasize some of basic properties of
  this $DB$ category, in order to render more selfcontained this
  presentation.\\
Every object  (denoted by $A,B,C$,..) of this category is a
  database instance, composed by a set of n-ary relations  $a_i\in A$, $i= 1,2,...$  called also "elements of
  $A$". \\
   In  ~\cite{Majk03} has been defined  the power-view operator $T$, with domain and codomain equal to the set of all database
  instances, such that for any object (database) $A$, the object
  $TA$  denotes a database composed by the set of \emph{all views} of $A$. The object $TA$, for a given database instance $A$, corresponds to
   the  quotient-term  algebra $\L_A/_\approx$,
  where carrier is a set of equivalence classes of closed terms of a
well defined formulae of a relational algebra, "constructed" by
$\Sigma_R$-constructors (relational operators in SPJRU algebra:
select, project, join and union) and symbols (attributes of
relations) of a database instance $A$, and constants of
attribute-domains.
  More precisely, $TA$ is "generated" by this quotient-term algebra
  $\L_A/_\approx$, i.e., for a given evaluation of queries in $\L_A$, $Eval_A:\L_A \rightarrow TA$, which is surjective
  function, from a factorization theorem, holds that there is a
  unique bijection $is_A:\L_A/_\approx \rightarrow TA$, such that the
  following diagram commutes
  \begin{diagram}
   \L_A  & \rTo^{Eval_{A}}   & TA\\
 \dTo^{nat_{\approx}} & \ruTo_{is_A} &\\
 \L_A/_\approx  &      &     \\
 \end{diagram}
 where the surjective function $nat_{\approx}:\L_A \rightarrow \L_A/_\approx$
 is a natural representation for the equivalence $\approx$.\\
  %
   For every object $A$ holds that $A \subseteq TA$, and $TA = TTA$, i.e., each (element) view
   of database instance $TA$ is also an element (view) of a database instance
   $A$.\\  \emph{Closed object} in $DB$ is a database $A$ such that $A = TA$. Notice that also when $A$ is  finitary
   (has a finite number of relations) but
   with at least one relation with infinite number of tuples, then $TA$ has an infinite number of relations (views of $A$),
    thus can be an infinitary object.
   It is obvious that when a domain of constants of a database is finite then both $A$ and $TA$ are finitary objects. \\
 From a \emph{behavioral point of view based on observations} we
can define equivalent (categorically isomorphic) objects (database
instances) as follows:  each arrow (morphism) is composed by a
number of "queries" (view-maps), and each query may be seen as an
\emph{observation} over some database instance (object of $DB$).
Thus, we can characterize each object in $DB$ (a database instance)
by its behavior according to a given set of observations.  Thus
databases $A$ and $B$ are equivalent (bisimilar) if they have the
same set of its observable internal states, i.e. when $TA$ is equal
to $TB$: $~~~~
A\approx B~~iff~~TA = TB $.\\
This equivalence relation corresponds to the isomorphism of objects
in $DB$ category \cite{Majk04AOT}. It is demonstrated that this powerview closure operator $T$ can be
extended also to arrows of $DB$ category, thus that it is an endofunctor and that \emph{defines a monad} (see Section 2). \\
 Basic properties of this
database category $DB$ as its symmetry (bijective correspondence
between arrows and objects, duality ($DB$ is equal to its dual
$DB^{OP}$) so that each limit is also colimit (ex. product is also
coproduct, pullback is also pushout, empty database $\bot^0$ is zero
objet, that is, both initial and terminal object, etc..), and that
it is a 2-category has been demonstrated in
\cite{Majk03,Majk04AOT}.\\
Generally, database mappings are not simply programs from values
(relations) into computations (views) but an equivalence of
computations: because of that each mapping, from any two databases A
and B, is symmetric and gives a duality property to the category
$DB$. The denotational semantics of database mappings is given by
morphisms of the Kleisli category $DB_T$ which may be "internalized"
in $DB$ category as "computations" \cite{MaBh10}.\\
 The product $A\times B$ of a databases
$A$ and $B$ is equal to their coproduct $A+B$, and the semantics for
them is that we are not able to define a view by using relations of
both databases, that is, these two databases have independent DBMS
for query evaluation. For example, the creation of exact copy of a
database $A$ in another DB server corresponds to the database $A +
A$. \\
In the paper \cite{Majk09a,Majk09f,Majk09a} have been considered
some relationships of $DB$ and standard $Set$ category, and has been
introduced the categorial (functors) semantics for two basic
database operations: \emph{matching} $\otimes$,  and \emph{merging}
$\oplus$, such that for any two databases $A$ and $B$, we have that
$A \otimes B = TA \bigcap TB$ and $A \oplus B = T(A \bigcup B)$. In
the same work has been defined the algebraic database lattice and
has been shown that $DB$ is concrete, small and locally finitely
presentable (lfp) category. Moreover, it was shown that $DB$ is also
V-category enriched over itself, was developed a metric space and a
subobject classifier for this
category, and demonstrated that it is a weak monoidal topos.\\
In this paper we will develop the denotational semantics for
database mappings based on power-view endofunctor $T$,
monadic $T$-(co)algebras and their computational properties in $DB$ category, and Kleisly category
of a monad $T$ used for categorial semantics of database mappings and database queries.\\
 Plan of this paper is the following: After
brief introduction of $DB$ category and its power-view monad $T$,
taken from \cite{Majk03,Majk04AOT,Majk08db}, in Section 3 consider
its behavioral equivalence and category symmetry. In Section 4 we
will consider universal algebra theory for databases and monadic
coalgebras for database mappings. In Section 5 will be developed the
categorial semantics of database mappings, based on Kleisly category
of the monad $T$. Finally in Section 6 are developed the theoretical
considerations of (co)algebras and (co)inductions for databases.

\section{Monad over DB Category}
In this section we will present a short introduction for a $DB$
category, based on work in \cite{Majk03,Majk04AOT,Majk08db}. As
default we assume that a domain of every database is arbitrary large
set but is
   finite. It is reasonable assumption for real applications.
    We define an universal database instance  $\Upsilon$,
  as the  union of all database instances, i.e., $\Upsilon = \{ a_i | a_i\in A, A\in Ob_{DB}\}$.
  It  is a top object of this category. We have that $\Upsilon = T\Upsilon$, because every view $v\in T\Upsilon$
  is  a database instance also, thus $v\in \Upsilon$; and vice versa, every element
   $r\in \Upsilon$ is also a view of $\Upsilon$, thus $r\in T\Upsilon$.\\
      Every object (database) $A$ has also an empty relation $\bot$. The object (database) composed by only this
  empty relation is denoted by $\bot^0$ and we have that $T\bot^0
  =\bot^0= \{\bot\}$. Any empty database (a database with only empty relations) is isomorphic to this bottom object $\bot^0$.\\
   Morphisms of this category are all possible mappings
  between database instances \emph{based on views}. Elementary view-map for a given database $A$ is given by a SPCU query $f_i = q_{A_i}:A \rightarrow
  TA$.
    Let us denote by $\|f_i\|$ the extension of the relation obtained by this query $q_{A_i}$. Suppose that $r_{i1},...,r_{ik} \in A$ are
the relations used for computation of this query, and that  the
corespondent algebraic term $~\widehat{q_i}$ is a function (it is
not a T-coalgebra) $\widehat{q_i}:A^k \rightarrow TA$, where $A^k$
is k-th cartesian product of $A$. Then, $\|q_{A_i}\| =
\widehat{q_i}(r_{i1},...,r_{ik})$. Differently from this algebra
term  $~\widehat{q_i}$ which is a function, a view-map $q_{A_i}:A
\rightarrow  TA$, which is \emph{a T-coalgebra}, \emph{is not} a function.\\
Consequently, an atomic morphism $f:A \rightarrow B$, from a
database $A$ to
database $B$, is a set of such view-mappings, thus \emph{it is not} generally a function.\\
We can introduce two functions, $\partial_0, \partial_1:Mor_{DB}
\rightarrow \P(\Upsilon)$ (which are different from standard
category functions $dom, cod:Mor_{DB} \rightarrow Ob_{DB}$), such
that for any view-map $q_{A_i}:A \longrightarrow TA$, we have that
   $\partial_0(q_{A_i}) = \{ r_1,..., r_k\} \subseteq A$  is a subset of relations of $A$
    used as arguments by this query $q_{A_i}$ and
   $\partial_1(q_{A_i})= \{v\}, v \in TA$ ($v$ is a resulting view of a query
   $q_{A_i}$). In fact, we have that they are functions $~\partial_0,
\partial_1:Mor_{DB} \rightarrow \P(\Upsilon)$ (where $\P$ is a powerset
operation), such that for any morphism $f:A \rightarrow B$ between
databases $A$ and $B$, which is a set of view-mappings $q_{A_i}$
such that $\|q_{A_i}\| \in B$, we have that $\partial_0(f) \subseteq
A$ and $\partial_1(f) \subseteq TA \bigcap B \subseteq B$.
  Thus,  we have
   \[
\partial_0(f)=  \bigcup_{ q_{A_i}\in f}
\partial_0 (q_{A_i}) \subseteq dom(f) = A,~~
~~ \partial_1 (f) = \bigcup_{ q_{A_i}\in f}
\partial_1 (q_{A_i})\subseteq cod(f) = B~~
\]
Based on atomic morphisms (sets of view-mappings) which are complete
arrows (c-arrows), we obtain that their composition generates
tree-structures, which can be incomplete (p-arrows), in the way that
for a composed arrow $h = g\circ f:A \rightarrow C$, of two atomic
arrows $f:A \rightarrow B$ and $g:B\rightarrow C$, we can have the
situations where $\partial_0(f) \subset
\partial_0(h)$, where the set of relations in $\partial_0(h) -
\partial_0(f) \subset
\partial_0(g)$ are denominated "hidden elements".
\begin{definition}\label{def:morphisms}  The following BNF defines the set $Mor_{DB}$ of all morphisms in
DB:\\
  $~~~p-arrow \textbf~{:=} ~c-arrow ~|~c-arrow \circ c-arrow~$ (for any two c-arrows  $f:A\longrightarrow B$ and $g:B\longrightarrow C~$)\\
 $~~~morphism \textbf~{:=} ~p-arrow ~|~c-arrow \circ p-arrow~$ (for any p-arrow $f:A\longrightarrow B$ and c-arrow $g:B\longrightarrow C$)\\
\\whereby the composition of two arrows, f (partial) and g (complete),
we obtain the following p-arrow (partial arrow) $h = g \circ
f:A\longrightarrow C$
\[
\ h = g\circ f = \bigcup_{ q_{B_j}\in ~g ~\& ~\partial_0 (q_{B_j})
\bigcap
\partial_1 (f) \neq \emptyset } \{q_{B_j}\}~~~~\circ
\]
\[
\circ ~~~~ \bigcup_{q_{A_i}\in ~f ~\&~\partial_1 (q_{A_i})=
\{v\}~\&~ v \in ~\partial_0 (q_{B_j}) } \{q_{A_i}(tree)\}~~
\]
$= \{q_{B_j} \circ \{q_{A_i}(tree)~|~\partial_1(q_{A_i}) \subseteq
\partial_0(q_{B_j})\}~|~q_{B_j}\in ~g ~\&
~\partial_0 (q_{B_j}) \bigcap
\partial_1 (f) \neq \emptyset \}$\\
$= \{q_{B_j}(tree)~|~q_{B_j}\in ~g ~\& ~\partial_0 (q_{B_j}) \bigcap
\partial_1 (f) \neq \emptyset\}$
\\where   $q_{A_i}(tree)$ is the  tree of the morphisms f below
$q_{A_i}$.\\
We define the semantics of mappings by function
 $B_T:Mor_{DB}\longrightarrow Ob_{DB}$, which, given any mapping
 morphism
 $f:A\longrightarrow B~$ returns with the set of views ("information flux") which are
 really "transmitted" from the source to the target object.\\
 1. for atomic morphism, $ \widetilde{f} = B_T(f)~\triangleq
 T\{\|f_i\|~|~f_i \in f \}$.\\
 2. Let $g:A \rightarrow B$ be a morphism with a flux
 $\widetilde{g}$, and $f:B \rightarrow C$ an atomic morphism with
 flux $\widetilde{f}$ defined in point 1, then  $~~\widetilde{f \circ g}
 = B_T(f \circ g) ~\triangleq \widetilde{f}\bigcap
 \widetilde{g}$.\\
 We introduce an equivalence relation over
morphisms by, $ ~~~~f \approx g ~~~~iff~~~~ \widetilde{f} =
\widetilde{g}$.
 \end{definition}
Notice that between any two databases $A$ and $B$ there is at least
an "empty" arrow $\emptyset:A \rightarrow B$ such that
$\partial_0(\emptyset) =
\partial_1(\emptyset) = \widetilde{\emptyset} = \{\bot\} = \bot^0$.
We have that $\bot \in A$ for any database $A$ ( in $DB$ all objects
are pointed by $\bot$ databases), so that any arrow $f:A \rightarrow
B$ has a component empty mapping $\emptyset$ (thus also arrows are
pointed by $\emptyset$).
 Thus we have the following fundamental properties:
\begin{propo} \cite{Majk08db} \label{prop-morphism} Any mapping  morphism
 $f:A\longrightarrow B~$  is a closed object in DB, i.e., $~~\widetilde{f} =
T\widetilde{f}$, such that $\widetilde{f} \subseteq TA \bigcap TB$, and\\
  1. each arrow  such that
$~\widetilde{f} =   TB$ is an epimorphism $f:A\twoheadrightarrow B$,\\
  2. each arrow  such that $~\widetilde{f}=
  TA$ is a monomorphism $f:A\hookrightarrow B$,\\
  3. each monic and epic arrow is an isomorphism.
\end{propo}
If $f$ is epic then $TA \supseteq TB$;  if it is monic then $TA
\subseteq TB$. Thus we have an isomorphism of two objects
(databases), $~~A \simeq B$ iff $TA = TB$. \\
We define an ordering $\preceq$ between databases by $~~A \preceq
B~~$ iff $~~TA \subseteq TB$.\\ Thus, for any database $A$ we have
that $A\simeq TA$, i.e., there is an isomorphic arrow $is_A =
\{q_{A_i}~|~\partial_0(q_{A_i}) =
\partial_1(q_{A_i}) = \{v\}$ and $v \in A \}:A \rightarrow TA$ and
its inverse $is_A^{inv} = \{q_{TA_i}~|~\partial_0(q_{TA_i}) =
\partial_1(q_{TA_i}) = \{v\}$ and $v \in A \subseteq TA\}:TA
\rightarrow A$, such that their flux
is $\widetilde{is_A} = \widetilde{is_A^{inv}} = TA$.\\
The following duality theorem tells that, for any commutative
diagram in $DB$ there is also the same commutative diagram composed
by the equal objects and inverted equivalent arrows: This
"bidirectional" mappings property of $DB$ is a consequence of the
fact that the composition of arrows is semantically based on the
set-intersection commutativity property for "information fluxes" of
its arrows. Thus \emph{any limit diagram} in $DB$ has also its
\emph{"reversed" equivalent colimit diagram} with equal objects,
\emph{any universal property }has also its \emph{equivalent
couniversal property }in $DB$.
\begin{theo} \cite{Majk08db} \label{th:duality} there exists the controvariant functor
$\underline{S}= (\underline{S}^0,\underline{S}^1):DB \longrightarrow
DB$ such that
\begin{enumerate}
  \item $\underline{S}^0$ is the identity function on objects.
  \item for any arrow in DB, $f:A\longrightarrow B$ we have $\underline{S}^1(f):B\longrightarrow
  A$, such that $\underline{S}^1(f) \triangleq f^{inv}$, where $f^{inv}$
  is (equivalent) reversed morphism of $~f~~~$ (i.e., $\widetilde{f^{inv}}= \widetilde{f}$),\\
  $f^{inv}= is^{-1}_A \circ(Tf)^{inv}\circ is_B~$  with
\[
  \ (Tf)^{inv} ~\triangleq \bigcup_{ \partial_0(q_{TB_j})
   = \partial_1(q_{TB_j}) =\{v\} ~\& ~v \in~  \widetilde{f}} \{q_{TB_j}:TB
   \rightarrow TA \}
\]
  \item The category DB is equal to its dual category $DB^{OP}$.
\end{enumerate}
\end{theo}
%
Let us extend the notion of the type operator $T$ into the notion of
the endofunctor in $DB$ category:
\begin{theo} \cite{Majk08db} \label{th:endofunctor}  There exists the endofunctor $T =
(T^0,T^1):DB \longrightarrow DB$, such that
\begin{enumerate}
  \item for any object A, the object component $T^0$ is
  equal to the type operator T, i.e.,  $~~~~T^0(A)\triangleq TA$
  \item for any morphism $~f:A\longrightarrow B$, the arrow
  component $T^1$ is  defined by
  \[
  \ T(f) \triangleq T^1(f) =  \bigcup_{ \partial_0(q_{TA_i})
  =\partial_1(q_{TA_i}) =\{v\} ~\& ~v \in ~ \widetilde{f}} \{q_{TA_i}:TA
  \rightarrow TB\}
  \]
  \item Endofunctor T preserves properties of arrows, i.e., if a
  morphism $f$ has a property P (monic, epic, isomorphic), then
  also $T(f)$ has the same property: let $P_{mono}, P_{epi}
  ~and~\\P_{iso}$ are monomorphic, epimorphic and isomorphic
  properties respectively, then the following formula is true\\
  $\forall(f\in Mor_{DB})(P_{mono}(f)\equiv P_{mono}(Tf)$ and $P_{epi}(f)\equiv
  P_{epi}(Tf)$  and  $P_{iso}(f)\equiv P_{iso}(Tf)$.
  \end{enumerate}
\end{theo}
\textbf{Proof:} it can be found in \cite{Majk08db}
\\$\square$\\
The endofunctor $T$ is a right and left adjoint to identity functor
$I_{DB}$, i.e., $T \simeq I_{DB}$, thus we have for the equivalence
adjunction $< T, I_{DB}, \eta^C ,\eta >$ the unit $\eta^C:T\simeq
I_{DB}$ such that for any object $A$ the arrow $\eta^C_A \triangleq
\eta^C(A)\equiv is^{-1}_A:TA\longrightarrow A$, and the counit
$\eta:I_{DB} \simeq T$ such that for any $A$ the arrow $\eta_A
\triangleq \eta(A)\equiv is_A:A\longrightarrow TA$
are isomorphic arrows in $DB$ (By duality theorem holds $\eta^C = \eta^{inv}$).\\
The function $T^1:(A\longrightarrow
B)\longrightarrow(TA\longrightarrow TB)~$ is not higher-order
function (arrows in $DB$ are not functions): thus, there is no
correspondent monad-comprehension for the monad $T$, which
invalidates the thesis ~\cite{Wald90} that "monads $\equiv~$
monad-comprehensions". It is only valid that "monad-comprehension
$\Rightarrow~$ monads".\\
We have already seen that the views of some database can be seen as
its \emph{observable computations}: what wee need, to obtain an
expressive power of computations in the category $DB$, are
categorial computational properties, as known, based on monads:
\begin{propo} \label{prop:monad} The power-view closure 2-endofunctor $T =
(T^0,T^1):DB \longrightarrow DB~$ defines the monad $(T,\eta ,\mu)~$
and
  the comonad $(T,\eta^C ,\mu^C)~$ in DB, such that $\eta:I_{DB}\backsimeq
  T~$ and $\eta^C:T \backsimeq I_{DB}~$ are natural isomorphisms,
  while $\mu:TT \longrightarrow T~$ and $\mu^C:T \longrightarrow TT~$ are
  equal to the natural identity transformation $id_T:T \longrightarrow
  T~$ (because T = TT).
\end{propo}
\textbf{Proof:} It is easy to verify that all commutative diagrams
of the monad ($\mu_A \circ \mu_{TA} = \mu_A \circ T\mu_A~$, $\mu_A
\circ \eta_{TA} = id_{TA} = \mu_A \circ T\eta_A$) and the comonad
are diagrams composed by identity arrows. Notice that by duality we
obtain $\eta_{TA} = T\eta_A = \mu^{inv}_A$.
\\$\square$
\section{Categorial symmetry and behavioral equivalence}
Let us  now consider the problem of how to define equivalent
(categorically isomorphic) objects (database instances) from a
\emph{behavioral point of view based on observations}: as we see,
each arrow (morphism) is composed by a number of "queries"
(view-maps), and each query may be seen as an \emph{observation}
over some database instance (object of $DB$). Thus, we can
characterize each object in $DB$ (a database instance) by its
behavior according to a given set of observations. Indeed, if one
object $A$ is considered as a black-box, the object $TA$ is only the
set of all observations on $A$. So, given two objects $A$ and $B$,
we are able to define the relation of equivalence between them based
on the notion of the bisimulation relation. If the observations
(resulting views of queries) of $A$ and $B$ are always equal,
independent of their particular internal structure, then
they look equivalent to an observer. \\
In fact, any database can be seen as a system with a number of
internal states that can be observed by using query operators (i.e,
programs without side-effects). Thus, databases $A$ and $B$ are
equivalent (bisimilar) if they have the same set of observations,
i.e. when $TA$ is equal to $TB$:
\begin{definition}\label{def:strong-eq}
The relation of (strong) behavioral equivalence $ '\approx'$ between
objects (databases) in $DB$ is defined by
\[ \ A \approx B~~iff~~TA = TB \]
  the equivalence relation for morphisms is given by,
$~~~f \approx g ~~~~iff~~~~ \widetilde{f} = \widetilde{g}$.
\end{definition}
 This relation of behavioral equivalence between objects corresponds to the notion of
isomorphism in the category $DB$ (see  Proposition \ref{prop-morphism} ).\\
This introduced  equivalence relation for arrows $\approx$, may be
given  by an (interpretation) function $B_T:Mor_{DB}\longrightarrow
Ob_{DB}~$ (see Definition \ref{def:morphisms}), such that $\approx$
is equal to the kernel of $B_T$, ($\approx~ =~ kerB_T$), i.e., this
is a fundamental concept for  categorial symmetry ~\cite{Majk98}:
\begin{definition}\label{def:symmetry} \textsc{Categorial symmetry}:\\
Let C be a category  with an \textsl{equivalence} relation $~\approx
~\subseteq Mor_C \times Mor_C$
 for its arrows (equivalence
relation for objects is the isomorphism $~\backsimeq ~\subseteq Ob_C
\times Ob_C$) such that there exists a bijection between equivalence
classes of $\approx$ and $\backsimeq$, so that it is possible to
define a skeletal category $|C|$ whose objects are defined by the
imagine of a function $B_T:Mor_C \longrightarrow Ob_C$ with the
kernel  $~kerB_T = ~ \thickapprox$, and to define an associative
composition operator for objects $*$,  for any fitted pair
$g\circ f$ of arrows, by $~~ B_T(g) * B_T(f)= B_T(g\circ f)$.\\
For any arrow in C, $f:A\longrightarrow B$, the object $B_T(f)$ in
C, denoted by $~\widetilde{f}$, is denominated as a
\textsl{conceptualized} object.
 \end{definition}
 \textbf{Remark:} This symmetry property allows us to consider all the properties
 of an arrow (up to the equivalence) as properties of objects and their
composition as well. Notice that any two arrows are \textsl{equal}
if and only if they are equivalent and have the same source and the
target objects.\\
We have that in symmetric categories holds that $f \approx g$ iff
$\widetilde{f} \simeq \widetilde{g}$.\\
 Let
us introduce, for a category $C$ and its arrow category $C\downarrow
C$, an encapsulation operator $J:Mor_C\longrightarrow
Ob_{C\downarrow C}$, that is,
 a one-to-one function such that for any arrow $f:A\longrightarrow B$, $J(f) = <A,B,f>$ is its correspondent object in $C\downarrow C$,
 with  its inverse $~\psi~ $ such that $\psi(<A,B,f>) = f$.\\ We
 denote by  $F_{st},S_{nd}:(C \downarrow
C)\longrightarrow C ~$  the first and the second comma functorial
projections (for any functor $F:C \rightarrow D$ between categories
$C$ and $D$, we denote by $F^0$ and $F^1$ its object and arrow
component), such that for any arrow $(k_1; k_2):<A,B,f>\rightarrow
<A',B',g>$ in $C\downarrow C$ (such that $k_2 \circ f = g \circ k_1$
in $C$), we have that $F_{st}^0(<A,B,f>) = A, F_{st}^1(k_1; k_2) =
k_1$ and
$S_{nd}^0(<A,B,f>) = B, S_{nd}^1(k_1; k_2) = k_2$.\\
We denote by $\blacktriangle :C\longrightarrow (C\downarrow C)~$ the
 diagonal functor, such that for any object $A$ in a category $C$, $\blacktriangle^0(A) =
 <A,A,id_A>$.\\
 An important subset of symmetric categories are Conceptually
 Closed and Extended symmetric categories, as follows:
\begin{definition}\label{def:symCat} \textsl{Conceptually closed} category is a symmetric category C with a functor
$T_e = (T_e^0,T_e^1): (C\downarrow C)\longrightarrow C$
such that  $T_e^0 = B_T \psi$, i.e., $B_T = T_e^0 J$,
 with a natural isomorphism $~\varphi:T_e \circ \blacktriangle~
\backsimeq~I_{C}$, where $I_C$ is an identity functor for $C$.\\
 C is an \textsl{extended symmetric} category if  holds also $~~~\tau^{-1}\bullet \tau = \psi$, for  vertical
 composition of natural transformations $\tau: F_{st} \longrightarrow
T_e ~$ and $ \tau^{-1}:T_e \longrightarrow S_{nd} $.
\end{definition}
Remark: it is easy to verify that in conceptually closed categories,
it holds that   any arrow $f$ is equivalent to an \emph{identity} arrow, that is, $f \approx id_{\widetilde{f}}$.\\
It is easy  to verify also that in extended symmetric categories the following holds:\\
$~~ \tau = (T_e^1 (\tau_I F_{st}^{0} ;\psi ))\bullet
 (\varphi^{-1}F_{st}^{0}) ,~~ \tau^{-1} = (\varphi^{-1} S_{nd}^{0}) \bullet (T_e^1 ( \psi ;\tau_I S_{nd}^{0}
 ))$,\\ where $\tau_I : I_{C}\longrightarrow
 I_{C}~$ is an identity natural transformation (for any object $A$ in $C$, $\tau_I(A)
 =id_A$).\\
 \textbf{Example:} The $Set$ is an extended symmetric
category: given any function $~f:A\longrightarrow B$ , the
conceptualized object  of this function is the graph of this
function (which is a set), $\widetilde{f}= B_T(f) = \{(x,f(x))~|~x
\in A\}$.\\
The equivalence $\approx$ on morphisms (arrows) is defined by: two
arrows $f$ and $g$ are equivalent, $f \approx g$, iff they have the
same graph.\\ The composition of objects $*$ is defined as
associative composition of binary relations (graphs), $B_T(g \circ
f) = \{(x,(g\circ f)(x))~|~x \in A\} =  \{(y,g(y))~|~y \in B\} \circ
\{(x,f(x))~|~x \in A\} = B_T(g) * B_T(f)$.\\
$Set$ is also conceptually closed by the functor $T_e$, such that
for any object $J(f) = <A,B,f>$, $T_e^0(J(f)) = B_T(f) =
\{(x,f(x))~|~x \in A\}$, and for any arrow
$(k_1;k_2):J(f)\rightarrow J(g)$, the component $T_e^1$ is defined
by: \\for any $(x,f(x)) \in T_e^0(J(f)), ~~T_e^1(k_1; k_2)(x,f(x)) =
(k_1(x), k_2(f(x)))$.\\ It is easy to verify the compositional
property for $T_e^1$, and that $T_e^1(id_A; id_B) =
id_{T_e^0(J(f))}$. For example, $Set$ is also an extended symmetric
category, such that for any object $J(f) = <A,B,f>$ in $Set
\downarrow Set$, we have that $\tau(J(f)):A \twoheadrightarrow
B_T(f)$ is an epimorphism, such that for any $x \in A$,
$~~\tau(J(f))(x) = (x, f(x))$, while $\tau^{-1}(J(f)):B_T(f)
\hookrightarrow B$ is a monomorphism such that for any $(x, f(x))
\in B_T(f), \\~~\tau^{-1}(J(f))(x, f(x)) = f(x)$.\\
Thus, each arrow in $Set$ is a composition of an epimorphism and a
monomorphism.
\\$\square$\\
 Now we are ready to present a
formal definition for the $DB$ category:
\begin{theo}\label{th:symmetry} The category DB is an extended symmetric category, closed by
the functor $T_e = (T_e^0,T_e^1): (C\downarrow C)\longrightarrow C$,
where $T_e^0 = B_T \psi~$ is the object component of this functor
such that for any arrow $f$ in DB, $T_e^0(J(f)) = \widetilde{f}$,
while its arrow component $~T_e^1$ is defined as follows: for any
arrow $(h_1;h_2):J(f)\longrightarrow J(g)~$ in $DB\downarrow
  DB$, such that $g\circ h_1 = h_2\circ f~$ in DB, holds
\[
\ T_e^1(h_1;h_2) =  \bigcup_{ \partial_0(q_{\widetilde{f}_i})
=\partial_1(q_{\widetilde{f}_i}) =\{v\} ~\& ~v \in ~ \widetilde{h_2
\circ f}} \{q_{\widetilde{f}_i}\}
\]
The associative composition operator for objects $*$, defined for
any fitted pair $g\circ f$ of arrows, is the set intersection
operator $\bigcap$. \\Thus, $~~ B_T(g)
* B_T(f) = \widetilde{g}\bigcap \widetilde{f} = \widetilde{g\circ f} = B_T(g\circ
f)$.
\end{theo}
\textbf{Proof:}  Each object $A$ has its identity (point-to-point)
morphism
$ id_A = \\ \bigcup_{ \partial_0(q_{A_i}) =\partial_1(q_{A_i})
=\{v\} ~\& ~v \in A} \{q_{A_i}\}$
and holds the associativity  $\widetilde{h\circ (g\circ f)}=
\widetilde{h} ~\bigcap ~\widetilde{(g\circ f)}\\ = \widetilde{h}
~\bigcap ~\widetilde{g }~ \bigcap~ \widetilde{f}= \widetilde{(h\circ
g)}~\bigcap ~ \widetilde{f} = \widetilde{(h\circ g)\circ f}$. They
have the same source and target object, thus $h\circ(g\circ f)=
(h\circ g)\circ f$. Thus, $DB$ is a category. It is easy to verify
that also $T_e~$ is a well defined functor. In fact, for any
identity arrow $(id_A;id_B):J(f) \longrightarrow J(f)$ it holds that
$T_e^1(id_A;id_B) =  \bigcup_{
\partial_0(q_{\widetilde{f}_i}) =\partial_1(q_{\widetilde{f}_i})
=\{v\} ~\& ~v \in ~ \widetilde{id_B \circ f}}
\{q_{\widetilde{f}_i}\} = id_{\widetilde{f}} $ is the identity arrow
of $\widetilde{f}$. For any two arrows
$(h_1;h_2):J(f)\longrightarrow J(g)$, $(l_1;l_2):J(g)\longrightarrow
J(k)$, it holds that $\overline{T_e^1(h_1;h_2)\circ T_e^1(l_1;l_2)}
= \widetilde{T_e^1(h_1;h_2)}\bigcap \widetilde{T_e^1(l_1;l_2)} =
T(\widetilde{l_2\circ g})\bigcap T(\widetilde{h_2\circ f})=
\widetilde{l_2}\bigcap \widetilde{g}\bigcap \widetilde{h_2}\bigcap
\widetilde{f} = (by~l_2\circ f = g\circ h_1)=\widetilde{l_2}\bigcap
\widetilde{g}\bigcap \widetilde{h_1}\bigcap \widetilde{h_2}\bigcap
\widetilde{f} = (by~l_2\circ f = g\circ h_1)= \widetilde{l_2}\bigcap
 \widetilde{h_2}\bigcap \widetilde{f} = \widetilde{l_2\circ h_2 \circ f}= T_e^1(l_1\circ h_1;l_2\circ h_2)$,
 finally, $T_e^1(h_1;h_2)\circ T_e^1(l_1;l_2)= T_e^1(l_1\circ h_1;l_2\circ
 h_2)$. For any identity arrow, it holds that $id_A$, $T_e^0J(id_A) = \widetilde{id_A}
 = TA \simeq A$ as well, thus,  an isomorphism $~\varphi:T_e \circ \blacktriangle~
\backsimeq~I_{DB}$ is valid.
\\$\square$\\
\textbf{Remark:} It is easy to verify (from $~~\tau^{-1}\bullet \tau
= \psi$) that for any given morphism $~f:A\longrightarrow B~$ in
$DB$, the arrow $f_{ep} = \tau (J(f)):A \twoheadrightarrow
\widetilde{f}~$ is an epimorphism, and the arrow $f_{in} =
\tau^{-1}(J(f)):\widetilde{f} \hookrightarrow B~$ is a monomorphism,
so that \emph{any morphism f in DB is a composition of an
epimorphism and  monomorphism} $ f = f_{in} \circ f_{ep}$, with the
intermediate object equal to its "information flux" $\widetilde{f}$,
and with $f \approx f_{in} \approx f_{ep}$.
\section{Databases: Universal algebra and monads}
The notion of a monad is one of the most general mathematical
notions. For instance, \emph{every} algebraic theory, that is, every
set of operations satisfying equational laws, can be seen as a monad
(which is also a \emph{monoid} in a category of endofunctors of a
given category: the "operation" $\mu$ being the associative
multiplication of this monoid and $\eta$ its unit). Thus monoid
laws of the monad do subsume all possible algebraic laws.\\
In order to explore universal algebra properties
\cite{Majk09f,Majk09a} for the category
 $DB$, where, generally, morphisms are not functions (this fact complicates a definition of mappings from its morphisms
 into homomorphisms of the category of $\Sigma_R$-algebras), we
 will use an equivalent to $DB$ "functional" category, denoted by
 $DB_{sk}$, such that  its arrows can be seen as total
 functions.
 \begin{propo}\label{prop-skeletal}
 Let us denote by $DB_{sk}$ the full skeletal
 subcategory of DB, composed by  closed objects only.\\
 Such a category is equivalent to the category DB, i.e., there
 exists an adjunction of a surjective functor $T_{sk}:DB\longrightarrow DB_{sk}$ and an
 inclusion functor $In_{sk}:DB_{sk}\longrightarrow DB$, where $In^0_{sk}$ and $In^1_{sk}$ are two identity
functions, such that
 $T_{sk}In_{sk} = Id_{DB_{sk}}$ and $In_{sk}T_{sk}\simeq
 Id_{DB}$.\\
  There exists the faithful forgetful functor
 $F_{sk}:DB_{sk}\longrightarrow Set $, and $F_{DB}= F_{sk}\circ T_{sk}:DB\longrightarrow
 Set$,  thus $~DB_{sk}~$ and $DB$ are  \textsl{concrete} categories.
  \end{propo}
\textbf{Proof:}   It can be found in \cite{Majk09a}. The skeletal
category $DB_{sk}$ has closed
  objects only, so, for any mapping  $f:A\rightarrow B$, we obtain the arrow
   $f_T = T_{sk}^1(f):TA\longrightarrow TB$ can be expressed in a
  following  "total" form such that $\partial_0(f_T) = T_{sk}^0(A) =
  TA$,
  \[
  \ f_T \triangleq  \bigcup_{ \partial_0(q_{TA_i})
  =\partial_1(q_{TA_i}) = \{v\}~\&~ v \in ~ \widetilde{f}}
  \{q_{TA_i}\}
    \bigcup_{ \partial_0(q_{TA_i}) = \{v\}~\&~ v \notin \widetilde{f}~\&~\partial_1(q_{TA_i}) = \perp^0 }
    \{q_{TA_i}\}
  \]
  so that $f_R = F_{sk}^1(f_T):TA \rightarrow TB$  (the component for objects $F_{sk}^1$ is an identity) is
 \emph{a function} in \emph{Set}, $f_R = F_{DB}^1(f)$, such
that for any $v \in TA$, $~~f_R(v) = v$ if $v \in \widetilde{f}$;
$\bot$ otherwise.
\\$\square$\\
 In a given inductive definition one defines a value of a function
 (in our example the endofunctor $T$) on all (algebraic)
 constructors (relational operators). What follows is based on the
 fundamental results of the Universal algebra ~\cite{Cohn65}.\\
 Let $\Sigma_R$ be a finitary signature (in the usual algebraic
 sense: a collection $F_\Sigma $ of function \emph{symbols} together with
 a function $~ar:F_\Sigma \longrightarrow N$ giving the finite arity
 of each function symbol) for a single-sorted (sort of relations) relational
 algebra.\\
 We can speak of $\Sigma_R$-equations and their satisfaction in a
$\Sigma_R$-algebra, obtaining the notion of a $(\Sigma_R,
E)$-algebra theory. In a special case, when $E$ is empty, we obtain
a purely syntax version of Universal algebra, where $\K$ is a
category of all $\Sigma_R$-algebras, and the
quotient-term algebras are simply term algebras. \\
An \emph{algebra} for the algebraic theory (type) $(\Sigma_R, E )$
is given by a set $X$, called the \emph{carrier} of the algebra,
together with interpretations for each of the function symbols in
$\Sigma_R$. A function symbol $f \in \Sigma_R$ of arity $k$ must be
interpreted by a function $\widehat{f}_X:X^k \longrightarrow X$.
Given this, a term containing $n$ distinct variables gives rise to a
function $X^n\longrightarrow X $ defined by induction on the
structure of the term. An algebra must also satisfy the equations
given in $E$ in the sense that equal \emph{terms} give rise to
identical functions (with obvious adjustments where the equated
terms do not contain exactly the same variables). A
\emph{homomorphism} of algebras from an algebra X to an algebra Y is
given by a function $ g:X\longrightarrow Y $ which commutes with
operations of the algebra $g(\widehat{f}_X(x_1,..,x_k)) =
~\widehat{f}_Y(g(x_1),..,g(x_k))$.
 This generates a variety category $\K$ of all relational algebras.
Consequently, there is a bifunctor $E:DB^{OP}_{sk}\times \K
\longrightarrow Set$ (where $Set$ is the category of sets), such
that for any database instance $A$ in $DB_{sk}$ there exists the
functor $~E(A,\_ ~):\K\longrightarrow  Set$ with an universal
element $(U(A), \varrho)$, where  $\varrho\in E(A,U(A))$ ,
 $\varrho:A \longrightarrow U(A)$ is an inclusion function and
 $U(A)$ is a free algebra over $A$ (quotient-term algebra generated by a carrier
 database instance $A$), such that for any function $f\in E(A,X)$
 there is a unique homomorphism $~h~$ from the free algebra $U(A)$ into an
 algebra $X$, with $~f = E(A,h)\circ \varrho$.\\
 From the so called "parameter theorem" we obtain that there exists:
\begin{itemize}
  \item  a unique universal functor $~U:DB_{sk} \longrightarrow \K~$ such
 that for any given database instance $A$ in $DB_{sk}$ it returns with the free
$\Sigma_R$-algebra  $U(A)$ (which is a quotient-term algebra, where
a carrier is a set of equivalence classes of closed terms of a well
defined formulae of a relational algebra, "constructed" by
$\Sigma_R$-constructors (relational operators: select, project, join
and union SPJRU) and symbols (attributes and relations) of a
database instance $A$, and constants of attribute-domains. An
alternative for $U(A)$ is given by considering  $A$ as a set of
variables rather than a set of constants, then we can consider
$U(A)$ as being a set of \emph{derived operations} of arity $A$ for
this theory. In either case the operations are interpreted
syntactically $\widehat{f}([t_1],...,[t_k]) = [f(t_1,...,t_k)]$,
where, as usual, brackets denote equivalence classes), while, for
any "functional" morphism (correspondent to the total function
$F^1_{sk}(f_T)$ in Set, $F_{sk}:DB_{sk} \longrightarrow Set$)
$~f_T:A\longrightarrow B$ in $DB_{sk}$ we obtain the homomorphism $
f_H = U^1(f_T)$ from the $\Sigma_R$-algebra $U(A)$ into the
$\Sigma_R$-algebra $U(B)$, such that for any term $\rho (a_1,..,a_n)
\in U(A)$, $\rho \in \Sigma_R$, we obtain $f_H(\rho (a_1,..,a_n)) =
\rho (f_H(a_1),...,f_H(a_n))$, so, $f_H$ is an identity function for
algebraic operators and it is equal to the function $F^1_{sk}(f_T)$
for constants.
  \item its adjoint forgetful functor $F:\K \longrightarrow DB_{sk}$,
  such that for any free algebra $U(A)$ in $\K$ the object $F \circ U(A)$
  in $DB_{sk}$ is equal to its carrier-set $A$ (each term $\rho (a_1,...,a_n) \in U(A)$
   is evaluated into a view of this closed object $A$ in $DB_{sk}$)
  and for each arrow $U^1(f_T)$ holds
  that $F^1U^1(f_T) = f_T$, i.e., we have that $FU = Id_{DB_{sk}}$ and $UF =
  Id_{\K}$.
  \end{itemize}
 Consequently,  $U(A)$ is a quotient-term algebra,
where carrier is a set of equivalence classes of closed terms of a
well defined formulae of a relational algebra, "constructed" by
$\Sigma_R$-constructors (relational operators in SPJRU algebra:
select, project, join and union) and symbols (attributes of
relations) of a
database instance $A$, and constants of attribute-domains.\\
It is immediate from the universal property that the map $A \mapsto
U(A)$ extends to the endofunctor $F \circ U:DB_{sk}\longrightarrow
DB_{sk}$. This functor carries \emph{monad structure} $(F \circ U,
\eta, \mu)$ with $F \circ U$ an equivalent version of $T$ but for
this skeletal database category $DB_{sk}$. The natural
  transformation $\eta $ is given by the obvious "inclusion" of
  $A$ into $F\circ U(A): a\longrightarrow [a]$ (each view $a$ in an closed object $A$
   is an equivalence class of all algebra terms which produce this view).
    Notice that the natural transformation $\eta$ is the unit of this adjunction of $U$ and $F$,
    and that it corresponds to an inclusion function in $Set$, $\varrho:A \longrightarrow U(A)$, given above. The
  interpretation of $ \mu $ is almost equally simple. An element
  of $(F\circ U)^2(A) $ is an equivalence class of terms built up
  from elements of $F\circ U(A)$, so that instead of $t(x_1,...,x_k)$, a typical element of $(F\circ U)^2(A) $ is
  given by the equivalence class of a term $ t([ t_1 ],...,[ t_k ])$.
  The transformation $\mu $ is defined by map $ [t([ t_1 ],...,[ t_k ])]~\mapsto~ [t(t_1 ,.., t_k )]$. This make sense because a
  substitution of provably equal expressions into the same term
  results in provably equal terms.
\section{Database mappings and monadic coalgebras}

We will use monads ~\cite{McLn71,LaSc86,KePo93} for giving
\emph{denotational semantics to database mappings}, and more
specifically as a way of modeling computational/collection types
~\cite{Mogg89,Mogg91,BNTW95,PlPo01}: to interpret a database
mappings (morphisms) in the category $DB$, we distinguish the object
$A$ (database instance of type $A$) from the object $TA$ of
observations (computations of type $A$ without side-effects), and
take as a denotation of (view) mappings the elements of $TA$ (which
are view of (type) $A$). In particular, we identify the type $A$
with the object of values (of type $A$) and obtain the object of
observations by applying the unary type-constructor $T$ (power-view
operator) to
$A$.\\
 It is well known that each endofunctor
defines algebras and coalgebras (the left and right commutative
diagrams)
\begin{diagram}
TA     & \rTo^{Tf} & TB    &   & TA   & \rTo^{Tf_1} & TB  \\
\dTo^h &           &\dTo_k &   &\uTo^{h_1}&           & \uTo_{k_1} \\
A      & \rTo^{f}  & B     &   &  A   & \rTo^{f_1}  &  B  \\
\end{diagram}
We will use the following well-known definitions in the category
theory (the set of all arrows in a category $\M~$ from $A$ to $B$ is
denoted by $\M(A,B)$):
\begin{definition} The categories $~CT_{alg}~$ of T-algebras,
$~CT_{coalg}~$ of T-coalgebras, derived from an endofunctor $T$, are
defined ~\cite{AsLo91} as follows :
\begin{enumerate}
  \item the objects of $~CT_{alg}~$ are pairs (A,h) with $A\in
  Ob_{DB}$ and $h\in DB(TA,A)$ ; the arrows between objects (A,h)
  and (B,k) are all arrows $f\in DB(A,B)$ such that $k \circ Tf = f \circ h: TA \longrightarrow B$.
  \item the objects of $~CT_{coalg}~$ are pairs (A,h) with $A\in
  Ob_{DB}$ and $h\in DB(A,TA)$ ; the arrows between objects (A,h)
  and (B,k) are all arrows $f\in DB(A,B)$ such that $Tf \circ h = k \circ f : A \longrightarrow TB$.
\end{enumerate}
\end{definition}
\begin{definition} \label{def:monadic} The \textbf{monadic} algebras/coalgebras, derived from a monad $(T, \eta , \mu)$, are
defined ~\cite{AsLo91,McLn71} as follows:
\begin{itemize}
  \item Each T-algebra $(A, h:TA \longrightarrow A)$, where $h$ is a "structure map", such that
  holds $~h \circ \mu_A = h \circ Th~$ and $h \circ \eta_A =id_A ~$
  is a monadic T-algebra. The category of all monadic algebras
  $~T_{alg}~$ is a full subcategory of $~CT_{alg}$.
  \item Each T-coalgebra $(A, k:A \longrightarrow TA)$, such that
  holds $~ Tk \circ k = \mu^C_A \circ k~$ and $\eta^C_A \circ k =
  id_A~$ is a monadic T-coalgebra. The category of all monadic coalgebras
  $~T_{coalg}~$ is a full subcategory of $~CT_{coalg}$.
  \end{itemize}
\end{definition}
Note: The monad $(T, \eta, \mu)$ given by commutative diagrams
\begin{diagram}
T^3A &\rTo^{T\mu_A} & T^2A  &  & TA& \rTo^{\eta_{TA}} & T^2A      & \lTo^{T\eta_A} & TA  \\
\dTo^{\mu_A} &   & \dTo_{\mu_A} &  &    & \rdTo_{id_A}    & \dTo_{\mu_A}& \ldTo_{id_A}    &    \\
T^2A &\rTo^{\mu_A}      & TA    &  &    &                 & TA        &                 &     \\
\end{diagram}
defines the adjunction $<F^T, G^T, \eta^T, \mu^T>:DB\longrightarrow
T_{alg}~$ such that $G^T \circ F^T = T :DB\longrightarrow DB$ , $
~\eta^T = \eta ~$, $ ~\epsilon^T = \eta^{inv} ~$ and $~\mu = G^T
\epsilon^T F^T$. The functors $F^T:DB\rightarrow T_{alg}$ and
$G^T:T_{alg}\rightarrow DB$ are defined as follows: for any object
(database) $A$, $F^T(A) = (A, \eta_A^{inv}:TA\simeq A)$, while
$G^T(A, \eta_A^{inv}:TA\simeq A) = TA$; for arrows $F^T$ and $G^T$
are identity functions.
%
\begin{definition} \label{def:KliesliTriple} Given a monad $(T, \eta, \mu)~$ over a category $\M$, we have ~\cite{McLn71}:
\begin{itemize}
  \item \textbf{Kleisli triple}  is a
triple $(T, \eta, -^*)$, where for $f:A \longrightarrow TB~$ we have
$f^*:TA \longrightarrow TB$, such that the following equations hold:
$~\eta^*_A = id_{TA}~,~~f^*\circ\eta_A = f~~, ~g^*\circ f^* =
(g^*\circ f)^*$, for
$~f:A\longrightarrow TB$ and $~g:B\longrightarrow TC$.\\
A Kleisli triple satisfies the \textbf{mono requirement} provided
$\eta_A~$ is monic for each object A.
  \item \textbf{Kleisli category} $\M_T$ has the same objects as $\M$
  category.  For any two objects A,B there is the bijection
  between arrows $\theta: \M(A,TB)\longrightarrow \M_T(A,B)$. For
  any two arrows $f:A\longrightarrow B,~g:B\longrightarrow C~$ in
  $\M_T$, their composition is defined by $~~ g \circ f \triangleq
  \theta(\mu_C\circ T\theta^{-1}(g)\circ \theta^{-1}(f))$.
\end{itemize}
\end{definition}
%
%
 The mono requirement for monad $(T,~\eta ,\mu)~$ ~\cite{Mogg91} is satisfied because
$\eta_A:A \longrightarrow TA~$ is a isomorphism $\eta_A = is_A$ (we
denote its inverse by $\eta^{-1}_A~$), thus it is also monic.
Consequently, the category $DB$ is  a \emph{computational model} for
view-mappings (which are programs) based on observations (i.e.,
views) with the typed operator $T$, so that:
\begin{itemize}
  \item $TA$ is a \emph{type of computations} (i.e. observations of the
  object of values  $A$ (of type $A$), which are the views of the database $A$)
  \item $\eta_A~$ is the \emph{inclusion} of values into computations
  (i.e., inclusion of elements of the database $A$ into the set of views
  of the database $A$). It is the isomorphism $\eta_A~ = is_A:A\longrightarrow TA$
  \item $f^*~$ is the \textbf{equivalent} \emph{extension} of a database mapping
  $f:A\longrightarrow TB~$ "from values to computations" (programs correspond to call-by-value parameter passing)
   to a  mapping "from computations to computations" (programs correspond to call-by-name),
   such that holds $ f^* = Tf = \mu_B \circ f \circ \eta^{-1}_A~$, so $~~f^*\approx
   f~$.\\ Thus, in $DB$ category, call-by-value ($f:A\longrightarrow TB$) and
   call-by-name ( $f^*:TA\longrightarrow TB$) paradigms of programs
   are represented by \textbf{equivalent} morphisms, $f\approx
   f^*$. Notice that in skeletal category $DB_{sk}$ (which is equivalent to $DB$)
   all morphisms correspond to the call-by-name paradigm, because
   each arrow is a mapping from computations into computations
   (which are closed objects).
\end{itemize}
The basic idea behind the semantic of programs ~\cite{Mogg89} is
that a program denotes a morphism from $A$ (the object of
\emph{values} of type $A$) to $TB$ (the object of
\emph{computations} of type $B$), according to the view of "programs
as functions from values to computations", so that the natural
category for interpreting programs (in our case a particular
equivalent "computation" database mappings of the form $f_1
\triangleq \eta_B\circ f : A \longrightarrow TB$, derived from a
database mapping $f:A\longrightarrow B$, such that $f_1 \approx f$)
is not $DB$ category, but the Kleisli category $DB_T$.\\
But, in our case, the Kleisli category is a perfect model only for a
subset of database mappings in $DB$: exactly for every view-mapping
(i.e., query) $~q_A:A\longrightarrow TA$ which is just an arrow in
Kleisli category $\theta(q_A):A \longrightarrow A$. For a general
database mapping $f:A\longrightarrow B$ in DB, only its (equivalent
to $f$) "computation extension" $\eta_B\circ f : A \longrightarrow
TB$ is an arrow $\theta(\eta_B\circ f):A \longrightarrow B$ in the
Kleisli category. Consequently, the Kleisli category is a model for
database mappings up to the equivalence "$\approx$".
\\It means that, generally, database mappings are not simply
programs from values into computations. In fact, the semantics of a
database mapping, between any two objects $A$ and $B$, is equal to
tell that for some set of computations (i.e, query-mappings) over
$A$ we have the same equivalent (in the sense that these programs
produce the same computed value (view)) set of computations
(query-mappings) over $B$: it is fundamentally an \emph{equivalence}
of computations. This is a consequence of the fact that each
database mapping (which \emph{is not a function}) from $A$ into $B$
is naturally bidirectional, i.e, it is a morphism
$f:A\longrightarrow B$ and its equivalent reversed morphism
$f^{inv}:B\longrightarrow A$ \emph{together} (explained by the
duality property $DB = DB^{OP}$ \cite{Majk04AOT}). Let us  define
  this equivalence formally:
\begin{definition} \label{def:mappings} Each database mapping $h:A\longrightarrow
B~$ is an equivalence of programs (epimorphisms), $h_A \triangleq
\tau (J(h)):A\twoheadrightarrow TH $ and $h_B \triangleq \tau^{-1}
(J(h))^{inv}:B\twoheadrightarrow TH $ ($\tau$ and $\tau^{-1}$ are
natural transformations of a categorial symmetry) , where $H$
generates a closed object $\widetilde{h}$ (i.e., $TH =
\widetilde{h}$ ) and $h_A \approx h \approx h_B$, such that
computations of these two programs (arrows of Kleisli category
$DB_T$ ) are equal, i.e., $~~~\partial_1(h_A) = \partial_1(h_B)$.
\end{definition}
 We can also give an alternative
model for equivalent computational extensions of database mappings
in $DB$ category:
\begin{propo}\label{prop:computation}
 Denotational semantics of  each mapping $f$, between any two database
instances $A$ and $B$, is given by the unique equivalent
"computation" arrow $f_1 \triangleq \eta_B\circ f~$ in $~T_{coalg}~$
from the monadic T-coalgebra $(A,\eta_A )$ into a cofree monadic
T-coalgebra $(TB, \mu^C_B )$, $~~~~f_1:(A,\eta_A
)\longrightarrow(TB, \mu^C_B )$; or , dually, by the unique
equivalent arrow $f^{inv}_1 \triangleq (\eta_B\circ f)^{inv} =
f^{inv}\circ \eta^{inv}_B$ from the free monadic T-algebra $(TB ,
\mu_B)$ into the monadic T-algebra $(A,\eta^{inv}_A)$.
 \end{propo}
\textbf{Proof:} In fact, holds $ \mu^C_B\circ f = Tf \circ \eta_A~$,
because $ \mu^C_B = id_{TB}~$,$~\widetilde{\mu^C_B} \bigcap
\widetilde{f} = TB \bigcap \widetilde{f} = \widetilde{f}~$ and $
\widetilde{Tf} \bigcap \widetilde{\eta_A} = \widetilde{Tf} \bigcap
TA = \widetilde{Tf} \bigcap TTA = \widetilde{Tf} = \widetilde{f}~$,
(because $\widetilde{f}~$ is a closed object).
\\$\square$\\
Note that each view-map (query) $q_A:A \longrightarrow TA$ is just
equal to its denotational semantics arrow in $~T_{coalg}~$,
$~~~~q_A:(A,\eta_A )\longrightarrow(TA, \mu^C_A )$.\\
 It is well known that for a Kleisli
category there exists an adjunction $<F_T , G_T , \eta_T , \mu_T>$
such that we obtain the same monad $(T, \eta , \mu)$ , such that $T
= G_T F_T , \mu = G_T \varepsilon_T F_T , \eta = \eta_T$. Let us see
now how the Kleisli category $DB_T$ is "internalized" into the $DB$
category.
\begin{propo} \label{prop:Kleisli} The Kleisli category $DB_T$ of the monad $(T, \eta ,
\mu)$ is isomorphic to $DB$ category, i.e., it may be "internalized"
in $DB$ by the faithful forgetful functor $K = (K^0,K^1):DB_T
\longrightarrow DB$, such that $K^0$ is an identity function and
$K^1 \triangleq \phi \theta^{-1} $, where , for any two objects $A$
and
$B$,\\
 $~~~~\theta :DB(A,TB) \simeq DB_T(A,B)~$ is Kleisli and \\
 $~~~~\phi:DB(A,TB)\simeq  DB(A,B)~$ , such that $\phi(\_) = \eta^{inv}_{cod(\_)}\circ \_$ \\
 is $DB$ category  bijection respectively.\\
 We can generalize a "representation" for the base $DB$ category
 (instead of usual Set category): a "representation" of functor
 $K$ is a pair $<\Upsilon , \varphi>$ , $\Upsilon$ is the total
 object and $\varphi:DB_T(\Upsilon ,\_)\simeq K$ is a natural
 isomorphism, where the functor $DB_T(\Upsilon ,\_):DB_T\longrightarrow
 DB$ defines "internalized" hom-sets in $DB_T$ , i.e., $DB^0_T(\Upsilon
 ,B)\triangleq TB^\Upsilon ~$, $DB^1_T(\Upsilon
 ,f)\triangleq id_\Upsilon \otimes Tf$.
\end{propo}
\textbf{Proof:} Let prove that $\phi$ is really bijection in $DB$.
For any program morphism $f:A\longrightarrow TB$ we obtain $\phi(f)
= \eta^{inv}_B \circ f :A\longrightarrow B $ and, viceversa, for any
$g:A\longrightarrow B$ its inverse $\phi^{-1}(g) \triangleq \eta_B
\circ g$ , thus , $\phi \phi^{-1}(g) = \phi (\eta_B \circ g) =
\eta^{inv}_B \circ (\eta_B \circ g) = (\eta^{inv}_B \circ \eta_B )
\circ g = id_B \circ g = g$ (because $\eta_B$ is an isomorphism),
i.e., $\phi \phi^{-1}$ is an identity function. Also $\phi^{-1}\phi
(f) = \phi^{-1}(\eta^{inv}_B \circ f) = \eta_B \circ (\eta^{inv}_B
\circ f) = ( \eta_B \circ \eta^{inv}_B ) \circ f = id_{TB}\circ f =
f $, i.e., $\phi^{-1}\phi $ is an identity function, thus $\phi $ is
a
bijection.\\
Let us demonstrate that $K$ is a functor: For any identity arrow
$id_T = \theta (\eta_A):A\longrightarrow A$ in $DB_T$ we obtain
$K^1(id_T) = \phi \theta^{-1}(\theta (\eta_A)) = \phi (\eta_A) =
\eta^{inv}_A \circ \eta_A = id_A$ (because  $\eta_A$ is an
isomorphism) . For any two arrows $g_T:B \longrightarrow C$ and
$f_T:A \longrightarrow B$ in Kleisli category,  we obtain, $K^1(g_T
\circ f_T) = K^1( \theta(\mu_C \circ T \theta^{-1}(g_T) \circ
\theta^{-1}(f_T))$ (from def. Kleisli category) $~ =
\phi\theta^{-1}( \theta(\mu_C \circ Tg \circ f))~$ (where $g
\triangleq \theta^{-1}(g_T):B \longrightarrow TC$ , $~f \triangleq
\theta^{-1}(f_T):A \longrightarrow TB$ ) $~= \phi(g \circ
\eta^{inv}_B \circ f)~$ (easy to verify in $DB$ that $ \mu_C \circ
Tg \circ f = g \circ \eta^{inv}_B \circ f$) $~ = \eta^{inv}_C \circ
g \circ \eta^{inv}_B \circ f = \phi(g)\circ \phi(f) = \phi
\theta^{-1}(\theta (g))\circ \phi \theta^{-1}(\theta (f)) = K^1
(\theta \theta^{-1}(g_T)) \circ K^1 (\theta \theta^{-1}(f_T)) =
K^1(g_T) \circ K^1(f_T)$.\\
Thus, each arrow $f_T:A\longrightarrow B$ in $DB_T$ is
"internalized" in $DB$ by its representation $f \triangleq K^1(f_T)
= \phi \theta^{-1}(f_T) = \eta^{inv}_B \circ
\theta^{-1}(f_T):A\longrightarrow B$ , where
$\theta^{-1}(f_T):A\longrightarrow TB$ is a program equivalent to
the database mapping $f:A\longrightarrow B$, i.e.,
$\theta^{-1}(f_T)\approx f$.\\
$K$ is faithful functor, in fact,  for any two arrows $f_T , h_T:A
\longrightarrow B$ in $DB_T$ , $K^1(f_T) = K^1(h_T)$ implies $f_T
= h_T$ :\\
from $K^1(f_T) = K^1(h_T)$ we obtain $\phi \theta^{-1}(f_T) = \phi
\theta^{-1}(h_T)$ , if we apply a bijection $\phi \theta^{-1}$ we
obtain $\phi \theta^{-1}\phi \theta^{-1}(f_T) = \phi \theta^{-1}\phi
\theta^{-1}(h_T)$ , i.e., $\theta \theta^{-1}(f_T) = \theta
\theta^{-1}(h_T)$ , i.e., $f_T = h_T~~$ ( $\theta
\theta^{-1}$ and $\phi \phi^{-1}$ are identity functions).\\
Let prove that $K$ is an isomorphism:  from the adjunction $<F_T ,
G_T , \eta_T , \mu_T>: DB\longrightarrow DB_T$ , where $F_T^0$ is
identity, $F^{-1}_T \triangleq \theta \phi^{-1}$, we obtain that
$F_T \circ K = I_{DB_T}$ and $K \circ F_T = I_{DB} $ , thus, the
functor $K$ is an isomorphism of $DB$ and Kleisli category $DB_T$.
\\$\square$\\
Remark: It is easy to verify that a natural isomorphism $\eta :
I_{DB}\longrightarrow T$ of the monad $(T, \eta , \mu)$ is equal to
the natural transformation $\eta:K \longrightarrow G_T$. (consider
that $G_T:DB_T \longrightarrow DB$ is defined by, $G_T^0 = T^0$ and
for any $f_T:A\longrightarrow B$ in $DB_T$, $G_T^1(f_T) \triangleq
\mu_B \circ T\theta^{-1}(f_T):TA \longrightarrow TB$).\\ Thus, the
functor $F_T$ has two different adjunctions: the universal
adjunction $<F_T , G_T , \eta_T , \mu_T>$ which gives the same monad
$(T, \eta , \mu)$ , and this particular (for $DB$ category only)
isomorphism's adjunction $<F_T , K , \eta_I , \mu_I>$  which gives
banal identity monad.
\\We are now ready  to define the semantics of queries in DB category
and the categorial definition of \emph{query equivalence}. This is
important in the context of the Database integration/exchange and
for the theory of \emph{query-rewriting} ~\cite{Hale00}.\\
 When we define a mapping (arrow, morphism) $f:A\longrightarrow B$
 between two databases $A$ and $B$, implicitly we define the
 "information flux" $~\widetilde{f}~$ , i.e, the set of views of $A$
 "transmitted" by this mapping into $B$.
 Thus, in the context of query-rewriting we consider only
 queries (i.e., view-maps) which resulting view (observation) belongs to the
 "information flux" of this mapping. Consequently, given any two
 queries, $q_{A_i}:A \longrightarrow TA $ and $q_{B_j}:B \longrightarrow TB $
 , they have to satisfy (w.r.t. query rewriting constraints) the
 condition $\partial_1(q_{A_i}) \in \widetilde{f} $ (the
 $\partial_1(q_{A_i})$ is just a resulting view of this query) and $\partial_1(q_{B_j}) \in \widetilde{f}
 $. So,  the well-rewritten query over $B$, $q_{B_j}:B \longrightarrow TB
 $, such that it is equivalent to the original query, i.e., $q_{B_j}\approx
 q_{A_i}$ , must satisfy the condition $\partial_1(q_{B_j})=
 \partial_1(q_{A_i}) \in \widetilde{f}$.\\
 Now we can give the denotational semantics for a query-rewriting
 in a data integration/exchange environment:
\begin{propo} \label{prop:rewriting}  Each database query is a (non monadic) T-coalgebra. Any
morphism between two T-coalgebras $f:(A,q_{A_i}) \longrightarrow
(B,q_{B_j})$ defines the semantics for relevant query-rewriting,
when $\partial_1(q_{A_i}) \in \widetilde{f} $.
\end{propo}
\textbf{Proof:} Consider the following commutative diagram, where
vertical arrows are T-coalgebras,
\begin{diagram}
 TA        & \rTo^{Tf} & TB\\
 \uTo^{q_{A_i}}&           & \uTo^{q_{B_i}}     \\
 A         & \rTo^{f}  &  B\\
\end{diagram}
The morphism between two T-coalgebras $f:(A,q_{A_i}) \longrightarrow
(B,q_{B_j})$ means that holds the commutativity $ q_{B_j}\circ f =
Tf \circ q_{A_i}:A \longrightarrow TB ~$, and from duality property
we obtain that $q_{B_i} = Tf \circ q_{A_i} \circ f^{inv}$.
%
Consequently, we have that for a given mapping $f:A \rightarrow B$
between databases $A$ and $B$,  every query $q_{A_i}$ such that
$\partial_1(q_{A_i}) \in \widetilde{f} $ (i.e., $\widetilde{q_{A_i}}
\subseteq \widetilde{f}$), we can have an equivalent rewritten query
$q_{B_i}$ over a data base $B$. In fact, we have that
$\widetilde{q_{B_i}} = \widetilde{Tf} \bigcap \widetilde{q_{A_i}}
\bigcap \widetilde{f^{inv}} = \widetilde{q_{A_i}}$, because of the
fact that $\widetilde{q_{A_i}} \subseteq \widetilde{f}$ and
$\widetilde{f^{inv}} = \widetilde{Tf} = \widetilde{f}$.\\
Thus $q_{B_j} \approx q_{A_i}$.
\\$\square$
\section{(Co)Algebras and (Co)Induction}
 Let us consider the following properties
for monadic algebras/coalgebras in $DB$:
\begin{propo}\label{prop:AlgCoalg}
The following properties for the monad $(T,~\eta, \mu)~$ and
  the comonad $(T,~\eta^C, \mu^C)~$ hold:
\begin{itemize}
  \item The categories $CT_{alg}~$ and $CT_{coalg}$, of the
  endofunctor $T:DB\longrightarrow DB$, are isomorphic ($CT_{coalg} = CT_{alg}^{OP}$),
   complete and cocomplete. The
  object $(\bot^0,~id_{\bot^0}:\bot^0\longrightarrow \bot^0~)$ is an
  initial T-algebra in $CT_{alg}~$ and a terminal T-coalgebra in
  $CT_{coalg}$.
  \item For each object $A$ in
  $DB$ category there exist the unique monadic T-algebra $(A,~ \eta^C_A:TA \longrightarrow A)~$ and the
   unique comonadic T-coalgebra $(A,~ \eta_A:A \longrightarrow  TA)$,
    $\eta^C_A = \eta^{inv}_A $ (i.e., $~\eta^C_A \approx \eta_A = is_A \approx id_A$).
  \item The free monadic T-algebra $(TA,~ \mu_A:T^2A\longrightarrow TA
  )~$ is dual (and equal) to the cofree monadic T-coalgebra $(TA, ~\mu^C_A:TA\longrightarrow T^2A
  )$, $\mu^C_A = \mu^{inv}_A $ (i.e., $~\mu^C_A = \mu_A = id_{TA}$).
  \item The Kleisli triple over the category $DB$ satisfies the mono
  requirement.
\end{itemize}
\end{propo}
\textbf{Proof:} Lets define the functor $F:T_{alg}\longrightarrow
T_{coalg}$, such that for any T-algebra $(A, h:TA\longrightarrow A)$
we obtain the dual T-coalgebra $F^0(A,h) =(A, h^{inv}:A
\longrightarrow TA)$, with a component $F^1$ for arrows  an identity
function; and the functor $F:T_{coalg}\longrightarrow T_{alg}$, such
that for any T-coalgebra $(A , k:A\longrightarrow TA)$ we obtain the
dual T-algebra $G^0(A,k) =(A, k^{inv}:TA \longrightarrow A)$, with a
component $G^1$ for arrows  an identity function. Thus holds $FG =
I_{T_{coalg}}$ and $GF = I_{T_{alg}}$. $T_{alg}~$ and $T_{coalg}~$
are complete and
cocomplete as the base $DB$ category ($T_{coalg} = T_{alg}^{OP}$ ).\\
The rest is easy to verify: each monadic T-algebra/coalgebra is an
isomorphism. The free monadic T-algebra and the cofree monadic
T-coalgebra are equal because $TA = T^2A $, thus, $\mu_A~ ,\mu^C_A~$
are identity arrows (by duality theorem).
\\$\square$\\
 As we can see, each monadic T-coalgebra is an \emph{equivalent reversed}
 arrow in $DB$  of some monadic T-algebra , and vice versa: the fundamental duality property of
 $DB$ introduces the equivalence of monadic T-algebras and monadic
 T-coalgebras, thus the equivalence of the dichotomy "\emph{construction} versus \emph{observation}" or duality between
 induction and coinduction principles ~\cite{JaRu97}.
\subsection{Algebras and induction}
 We have seen (from Universal algebra considerations) that there exists
 the unique universal functor $~U:DB_{sk} \longrightarrow \K~$ such
 that for any given database instance $A$ in $DB_{sk}$ returns with the free
$\Sigma_R$-algebra  $U(A)$ .\\
Its adjoint is the forgetful functor $F:\K \longrightarrow DB_{sk}$,
such that for any free algebra $U(A)$ in $\K$ the object $F \circ
U(A)$   in $DB_{sk}$ is equal to its carrier-set $A$ ( each term
$\rho (a_1,..,a_n) \in U(A)$    is evaluated into some view of this
closed object $A$ in $DB_{sk}$).\\
 It is immediate from the universal property that the map $A \mapsto
  U(A)$ extends to the endofunctor $F \circ U:DB_{sk}\longrightarrow DB_{sk}
  $. This functor carries monad structure: the natural
  transformation $\eta $ is given by the obvious "inclusion" of
  $A$ into $F\circ U(A): a\longrightarrow [a]$ .\\
  \emph{Finitariness}: In a locally finitely presentable (lfp) category every object can
  be given as the directed (or filtered) colimit of the
  finitely presentable (fp) objects. Hence, if the action of a monad
  preserves this particular kind of colimits, its action on any
  object will be determined by its action on the fp objects; such
  a monad is called finitary.\\
  Let verify that the power-view closure 2-endofunctor $T:DB
\longrightarrow DB$ is just a composition of functors described
above and that it is \emph{finitary} monad.
\begin{propo}\label{prop:EilenMoore} The power-view closure 2-endofunctor $T:DB
\longrightarrow DB$ is immediate from the universal property of
composed adjunction $<U T_{sk}, In_{sk}F, ~In_{sk}\eta_U T_{sk}\cdot
\eta_{sk} ,~ \varepsilon_U \cdot U \varepsilon_{sk} F
>:DB \longrightarrow \K$, i.e., $ ~~~~ T = In_{sk}F U T_{sk} \simeq Id_{DB}$. It is finitary.\\
The category $DB$ is equivalent to the (Eilenberg-Moore) category
$T_{alg}$ of all monadic T-algebras and is equivalent  to the
category $T_{coalg}$ of all monadic T-coalgebras .\\ Its equivalent
skeletal category $DB_{sk}$ is, instead, isomorphic to $T_{alg}$ and
$T_{coalg}$.
\end{propo}
\textbf{Proof:} For any object A in $DB$ holds $In_{sk}F U T_{sk}(A)
= In_{sk}T_{sk}(A) = TA$,
 and for any morphism $f:A\longrightarrow B$ in $DB$ holds  $In_{sk}F U T_{sk}(f) = In_{sk} K_{sk}(f)
 = In_{sk}(f_T) = Tf~ $ (where from Proposition \ref{prop-skeletal}, $f_T = T_{sk}^1(f)$, and $ \widetilde{f_T} = \widetilde{Tf} = \widetilde{f}$).
\\The adjunction -
equivalence $<T_{sk}, In_{sk}, \eta_{sk}, \varepsilon_{sk}>$ between
$DB$ and $DB_{sk}$ and the adjunction-isomorphism $<U, F, \eta_U,
\varepsilon_U>$ $DB_{sk}\simeq \K$, give the composed adjunction
$<UT_{sk}, In_{sk}F, In_{sk}\eta_UT_{sk} \cdot \eta_{sk},
\varepsilon_U \cdot U\varepsilon_{sk} F>:DB\longrightarrow \K$,
which
is an equivalence.\\
We have that $\K \simeq DB_{sk}$, and, from universal algebra
(Back's theorem) theory, $~\K \simeq T_{alg}$, thus $DB_{sk} \simeq
T_{alg}$. From this facte and the fact that $DB$ is equivalent to
$DB_{sk}$ we obtain that $DB$ is equivalent to $T_{alg}$. The
property for $T_{coalg}$ holds by duality.\\
 To understand the
finitary condition, consider the term algebra $U(A)$ over infinite
database (infinite set of relations) $A$. Since every operation
$\rho \in \Sigma_R$ can only take finitely many arguments, every
term $t \in U(A)$ can only contain finitely many variables from $A$;
and hence, instead of building the term algebra over the
\textbf{infinite} database $A$, we can also build the term algebras
over \emph{all finite subsets}(of relations) $A_0$  of $A$ and take
union of these: $~~U(A) = \bigcup \{ U(A_0)~|~A_0 \subseteq_\omega A
\}$. This result comes from Universal algebra because the closure
operator $T$ is \emph{algebraic } and $<\C, \subseteq >$, where $\C$
is a set of all closed objects in $DB$, is an algebraic
(complete+compact) lattice.
\\$\square$\\
The notion of T-algebra subsumes the notion of a $\Sigma_R$-algebra
($\Sigma_R$-algebras can be understood as algebras in which
operators (of the signature) are not subject to any law, i.e., with
empty set of equations). In particular, the monad $T$ freely
generated by a signature $\Sigma_R$ is such that $T_{alg}$ is
isomorphic to the category of $\Sigma_R$-algebras. Therefore, the
\textbf{syntax} of a programming language can be identified with
monad, the \emph{syntactical monad }$T$ freely
generated by the program constructors $\Sigma_R$.\\
 We illustrate the link between a single-sorted (sort is a relation)
$\Sigma_R$ algebra signature of a relational algebra operators
 and the T-algebras of the endofunctor $T$. The assumption that the
signature $\Sigma_R$ is finite is not essential for the
correspondence between models of $\Sigma_R$ and algebras of $T$. If
$\Sigma_R$ is infinite one can define $T$ via an infinite coproduct,
commonly written as
\[
\ \Sigma_R(A) = \biguplus_{\sigma \in \Sigma_R}A^{ar(\sigma)}.
\]
which is a more compact way of describing the category of
$\Sigma_R$-algebras is by taking this coproduct in $Set$ category
(disjoint union) $\biguplus_{\sigma \in \Sigma_R}A^{ar(\sigma)}$ ,
$~1\leq ar(\sigma_i)\leq N~, ~ i = 1,2,..,n$ , where the set $A^m$
is the m-fold product $A\times A\times ..\times A$;  that is, the
disjoint union of domains of the operations $\sigma \in \Sigma_R$ of
this "select-project-join +union" language (SPJRU language
~\cite{AbHV95}). More formally, for the signature $\Sigma_R$ we
define the endofunctor $\Sigma_R:Set\longrightarrow Set$, such that
for any object $B$, $~\Sigma_R(B) \triangleq \biguplus_{\sigma \in
\Sigma_R}B^{ar(\sigma)} $, and any arrow in $Set$ (a function) $f:B
\longrightarrow C$ , $~\Sigma_R(f) \triangleq \biguplus_{\sigma \in
\Sigma_R}f^{ar(\sigma)}$.\\
Thus, also for any object $A$ in $Set$ we have the endofunctor
$\Sigma_{R_A}:Set\longrightarrow Set$, such that for any object $B$
in $Set$ holds $\Sigma_{R_A}(B) = (\Sigma_R + A)(B)= A +\Sigma_R B
\triangleq A + \biguplus_{\sigma \in \Sigma_R}B^{ar(\sigma)}$, and
any arrow $f:B \longrightarrow C$, $~\Sigma_{R_A}(f) \triangleq id_A
+ \biguplus_{\sigma \in
\Sigma_R}f^{ar(\sigma)}$.\\
 Let $\omega$ be the
category of natural numbers with arrows
 $\leq:j\longrightarrow k$ which correspond to the total order
 relation $j\leq k$, i.e., $\omega = \{ 0 \rightarrow 1\rightarrow 2\rightarrow
 ....\}$.  An endofunctor $H:C \longrightarrow D$ is \emph{$\omega-cocontinuous$} if preserves the
 colimits of functors $J:\omega \longrightarrow C$, that is when $H
 ColimJ \simeq ColimHJ$ (the categories $C$ and $D$ are thus supposed
 to have these colimits). Notice that a functor $J:\omega \longrightarrow
 C$ is a diagram in $C$ of the form $\{ C_0 \rightarrow C_1\rightarrow C_2\rightarrow
 ....\}$. For $~\omega-cocontinuous$ endofunctors the construction
 of the \emph{initial algebra} is inductive ~\cite{SmPl82}.\\
 We define an \emph{iteratable} endofunctor $H$ of a category $\D$ if
for every object $X$ of $\D$ the endofunctor $H(\_) + X$ has an
initial algebra. It is well known that the signature endofunctor
$\Sigma_R$  in $Set$ category is $\omega$-cocontinuous and iteratable.\\
%
The initial algebra
 for a given set of terms with
variables in $A$, $\T A$, of the endofunctor $\Sigma_{R_A}= A +
\Sigma_R:Set\longrightarrow Set$ comes with an induction principle,
and since it is the coproduct $A + \Sigma_R\T A$ , we can rephrase
the principle as follows: For every $\Sigma_R$-algebra structure
$h:\Sigma_RB\longrightarrow B$  and every mapping
$f:A\longrightarrow B$ there exists a unique arrow $f_\#:\T
A\longrightarrow B$ such that the following diagram in $Set$
\begin{diagram}
A      & \rInto^{inl_A} &   \T A       & \lInto^{inr_A} & \Sigma_R \T A  \\
       & \rdTo^f        & \dTo_{f_\#}&                & \dTo_{\Sigma_Rf_\#}\\
       &                &   B        & \lTo^h         & \Sigma_R B \\
\end{diagram}
commutes, where $f_\# = [f , h \circ \Sigma_Rf_\#]$ is the unique
\emph{\textbf{inductive extension }of} $h$ \emph{along the mapping}
$f$. \\The arrow $inl_A:A \hookrightarrow \T A$ is an inclusion of
variables in $A$ into terms with variables $\T A$. Formally, $r_i
\in A$ is an element of a set $A$ of relations, and only after
applying $inl_A$ tom it that one obtain a variable. The arrow
$inr_A:\Sigma_R \T A \hookrightarrow \T A$ is an injection which
permits to construct a new term given any n-ary algebraic operator
$\sigma \in \Sigma_R$ and terms $t_1,..., t_n$ in $\T A$. Also the
right injection is usually left implicitly and one writes simply
$\sigma(t_1,..., t_n)$ for the resulting term.
 \\
Notice that $f_\#:<\T A, [inl_A , inr_A ]>\longrightarrow <B, [f , h
]>$ is the unique arrow from the initial
$\Sigma_{R_A}$-algebra to the algebra of the structure map $[f , h ]:\Sigma_R B + A \longrightarrow B$.\\
From Lambek's theorem, this initial $A + \Sigma_R$-algebra (that is,
the free $\Sigma_R$ algebra with carrier set $A$) is an isomorphism
$isa = [inl_A , inr_A ]:(A+\Sigma_R \T A) \simeq \T A$.\\\\
\textbf{Inductive principle in the $DB$ category:}\\
From the fact \cite{Majk09a} that $DB$ is a lfp category enriched
over  the lfp symmetric monoidal closed category with a tensor
product   $\otimes$ (matching operator for databases), and the fact
that $T$   is a finitary enriched monad on $DB$, by Kelly-Power
theorem we   have that $DB$ admits a presentation by operations and
equations,  so that $DB$ is the category of models for an
essentially algebraic  theory.\\
 Let us denote by $DB_I$ the "poset" subcategory of $DB$ with the same objects and with only monic arrow $in_B:B \hookrightarrow A$ iff $B\preceq A$
 (i.e., $TB \subseteq TA$).
  Than we can introduce
  a functor $\Sigma_D:DB_f\rightarrow DB$, where $DB_f$ is a full subcategory of $DB_I$ composed by
only \emph{finite objects} (databases), for the
  signature of relational algebra w.r.t. the lfp category $DB$ enriched over
  itself but where the arrows are not functions, analogously to standard algebra signature $\Sigma_R$
  defined over $Set$ category where arrows are funtions. This definition of $\Sigma_D$ is correct because all sigma
  operations $\sigma \in \Sigma_R$
  of relational algebra are finitary, i.e., with arity $n = ar(\sigma)$ a finite number, thus an arrow $f_\sigma$ in $DB$ which represents such an operation from a database $A$ into closed database $TA$
  will have finite cardinality of $\partial_0(f_\sigma)\subseteq_\omega A$, with cardinality $|\partial_0(f_\sigma)| = ar(\sigma)$,
   so that we can restrict $\Sigma_D$ to finite databases only.
   The extension of $\Sigma_D$ to all databases, as infinite databases which are not closed objects (i.e., compact objects in $DB$),
   can be successively obtained by  left Kan extension of this finite restriction as will be demonstrated in what follows. \\
First of all we have to demonstrate the existence of an
$\omega$-cocontinuous endofunctor for $DB$ category which can be
used for a construction of the initial algebra based on morphisms of
$DB$ category which are not functions as in the standard case of
$Set$ category.
\begin{propo} \label{prop:cocontinuity} For each object $A$ in the category $DB$ the "merging with $A$"
  endofunctor $\Sigma_A = A\oplus \_:DB\longrightarrow DB$, and  the endofunctor $A + T \_:DB\longrightarrow DB$
  are $\omega-cocontinuous$.
 \end{propo}
 \textbf{Proof:} Let us consider any chain in $DB$ (all arrows are
 monomorphisms, i.e., "$\preceq$" in a correspondent chain of the $<Ob_{DB}, \preceq >$ algebraic
 lattice), is a following diagram $J:\omega \longrightarrow DB$, \\
 $  \perp^0 ~\preceq_0 ~(\Sigma_A \perp^0) ~\preceq_1~ (\Sigma^2_A \perp^0)~ \preceq_2~~~
 ... ~~~~~~\Sigma^{\omega}_A  $,\\
  where $\perp^0$ is the initial object in $DB$, with unique monic arrow $\perp = \preceq_0:\perp^0 \hookrightarrow (\sum_A
  \perp^0)$ with $\widetilde{\perp} = \perp^0$,
  and consecutive arrows $\preceq_n = \Sigma^n_A \perp: (\Sigma^n_A \perp^0) \hookrightarrow
  (\Sigma^{n+1}_A \perp^0)$ with  $\widetilde{\Sigma^n_A \perp} = TA$, for all $n \geq 1$,
 as representation of a
 functor (diagram) $J:\omega \longrightarrow DB$. The endofunctor $\Sigma_A$
 preserves colimits because it is monotone and $\Sigma^{\omega}_A = TA$ is its
 fixed point, i.e., $\Sigma^{\omega}_A = TA  = T(A \bigcup TA) = T(A \bigcup \Sigma^{\omega}_A) = \Sigma_A(\Sigma^{\omega}_A)$.
 Thus, the colimit   $ColimJ = \Sigma^{\omega}_A$  of
 the base diagram $\D$ given by the functor $J:\omega \longrightarrow DB$, is equal to $ColimJ= (A\oplus \_)^{\omega}\perp^0 = TA$.
 Thus $\Sigma_AColimJ = T(A \bigcup ColimJ) = T(A\bigcup TA) = T(TA) = TA = Colim\Sigma_AJ$ (where $Colim\Sigma_AJ$ is a colimit of the diagram
 $\Sigma_A J$).\\
  The $\omega-cocompleteness$ amounts to chain-completeness,
 i.e., to the existence of least upper bound of $\omega-chains$.
 Thus $\Sigma_A$ is $\omega-cocontinuous$ endofunctor: a monotone
 function which preserves lubs of $\omega-chains$.\\
  Constant endofunctor $A:DB \rightarrow DB$ is $\omega$-cocontinuous
 endofunctor, identity endofunctors are $\omega$-cocontinuous,
 colimit functors (thus coproduct $+$) are $\omega$-cocontinuous
 (because of the standard "interchange of colinits"). Since
 $\omega$-cocontinuousness is preserved by functor composition
 $\circ$, then for the second endofunctor $A + T\_ = (A + Id \_)\circ T$ it is enough
 to show that $T$ is $\omega$-cocontinuous endofunctor. In fact
 consider the following diagram obtained by iterative application of
 the endofunctor $T$\\
$  \perp^0 ~\preceq_0 ~(T \perp^0) ~\preceq_1~ (T^2 T^2 \perp^0)~
\preceq_2~~~
 ... ~~~~~~T^{\omega}\perp^0  $,\\
  where $\perp^0$ is the initial object in $DB$,  and all objects $T^n \perp^0 =
  \perp^0$, so that all arrows in this chain are identities. Thus we
  obtain that $ColimJ = T^{\omega}\perp^0 = \perp^0$, and holds $TColimJ = ColimTJ = \perp^0$, so
  that $T$ is $\omega$-cocontinuous endofunctor.
 \\$\square$\\
In what follows we will make the translation of inductive principle
from $Set$ into $DB$ category, based on the following
considerations:
\begin{itemize}
  \item The object $A$ in $Set$ is considered as set of variables
  (for relations in a database instance $A$) while in $DB$ this
  object is considered as set of relations. Analogously, the set of
  terms with variables in $A$, $\T A$, used in $Set$ category, is
  translated into set $TA$ of all views (which are relations obtained by
  computation of these terms with variables in $A$).\\
  But it is not a carrier set for the initial $(A + \Sigma_D)$-algebra for $\Sigma_D = T$ (see below), just because generally $TA$ is not isomorphic
  to $A +\Sigma_D(TA) = A + TA$ (in fact $T(TA) = TA \neq T(A + TA)
  = TA +TA$).\\
 %
  \item Cartesian product $\times:Set \rightarrow Set$ is
translated into matching operation
(tensor product) $\otimes:DB\longrightarrow DB$.\\
This translation is based on observations that  any n-ary algebraic
operator $\sigma \in \Sigma_R$, is represented as an function
(arrow) $\sigma:\T A^n \rightarrow \T A$ which use as domain the
n-fold cartesian product $\T A \times ... \times \T A$, while such
an operator in $DB$ category is represented by  view-based mapping
$f_{\sigma} = \{q^{\sigma}_j~|~ \partial_0(q^{\sigma}_j)
=\{r_{i1},...,r_{in} \},
\partial_1(q^{\sigma}_j) = \{\sigma(r_{i1},...,r_{in}) \}$ for each
tuple $(r_{i1},...,r_{in}) \in A^n  \}:TA \rightarrow TA$.\\ Thus
this algebraic operator $\sigma$ is translated into an arrow from
$TA$ into $TA$. In fact if we replace $\times$ by $\otimes$ in
n-fold $\T A \times ... \times \T A$, we obtain $T A \otimes ...
\otimes TA = T(TA) \bigcap ... \bigcap T(TA) = T(TA) = TA$.\\
  \item Any disjoint union   $X + \_:Set\rightarrow Set$ used for construction of $\Sigma_R$
  endofunctor is translated into
"merging with X" endofunctor $ X\oplus
\_:DB\longrightarrow DB$. \\
From the fact that coproduct $+$ is replaced by merging operator
$\oplus$, we obtain that the object $\Sigma_R(X) = \biguplus_{\sigma
\in \Sigma_R}X^{ar(\sigma)}$ in $Set$ is translated by the object
$\Sigma_D (X) = \oplus_{\sigma \in \Sigma_R}(X\otimes ...\otimes X)
= \oplus_{\sigma \in \Sigma_R}TX = TX $, where the endofunctor
$\Sigma_D = T:DB_f \rightarrow DB$ is the translation for the
relational-algebra signature endofunctor $\Sigma_R:Set\rightarrow
Set$.
\end{itemize}
It is well known \cite{McLn71} that for any monoidal category $\A$
with a monoidal product $\otimes$,  any two functors $F_1:P^{op}
\rightarrow \A$ and $F_2:P \rightarrow \A$ have a tensor functorial
product $F_1 \bigotimes_{P} F_2 = \int^{p \in P}(F_1p)\otimes
(F_2p)$.\\
In our case we take for $\A$ the lfp \emph{enriched (co)complete}
category $DB$ with monoidal product correspondent to matching
database operation $\otimes$, $F_2 = \Sigma_D:DB_f \rightarrow DB$
and for $F_1$ the hom functor for a given database $A$ (object in
$DB$), $DB_I(\_,A)\circ K:DB_f \rightarrow DB$, where $K:DB_f
\hookrightarrow DB_I$ is an inclusion functor. Notice that for any
finite database, i.e., object $B \in DB_f$, the  $DB_I(K(B),A)$ is a
hom-object $A^{K(B)}$ of enriched database subcategory $DB_I$. In
this context we obtain that for any object (also infinite) $A$ in
$DB_I$ (that is , in $DB$), we have a tensor product
$DB_I(\_,A)\circ K \bigotimes_{P} \Sigma_D = \int^{B \in
DB_f}DB_I(K(B),A)\otimes \Sigma_D B = \int^{B \in
DB_f}DB_I(B,A)\otimes \Sigma_D B$.\\
This tensorial product comes with a dinatural transformation
\cite{DuSt70} $\beta:S \rightarrow A$, where $S = DB_I(\_,A)\otimes
\Sigma_D \_:DB_f^{OP}\times DB_f \rightarrow DB$ and $A$ is a
constant functor between the same categories of the functor $S$.
Thus, for any given object $A$ in $DB$ we have a collection of
arrows $\beta_B:DB_I(B,A)\otimes \Sigma_D B \rightarrow A$ (for
every object $B \in DB_f)$.
 \\
In the case of standard case of $Set$, which is (co)complete lfp
with monoidal product $\otimes$ equal to cartesian product $\times$,
we have that such arrows are $\beta_B:Set(B,A)\times \Sigma_R (B)
\rightarrow A$, where $B$ is a finite set with cardinality $n =
|B|$, so that $Set(B,A)$ is a set of all tuples of arity $n$
composed by elements of the set $A$, while $\Sigma_R (B)$  here is
\emph{interpreted} as a set of all \emph{basic n-ary} algebra
operations. So that $\beta_B$ is a specification for all basic
algebra operations with arity $n$, and is a funtion such that for
any n-ary operation $\sigma \in \Sigma_R(B)$ and a tuple
$<a_1,...,a_n> \in A^n \simeq Set(B,A)$, (where $\simeq$ is an
isomorphism in $Set$),
returns with result $~~\beta_B(<a_1,...,a_n>, \sigma) = \sigma(a_1,...,a_n) \in A$.\\
In the non standard case, when instead of base category $Set$ is
used another lfp enriched (co)complete category, as $DB$ category in
our case, the \emph{interpretation} for this tensorial product and
dinatuaral transformation $\beta$ is obviously very different, as we
will see in what follows.\\
 From considerations explained previously we obtain that the finitary signature functor
$\Sigma_D:DB_f \rightarrow DB$ has a left Kan extension
\cite{Dubu70} in enriched category $DB$  $Lan_K(\Sigma_D):DB_I
\rightarrow DB$ and left Kan extension $Lan_{J\circ K}(\Sigma_D):DB
\rightarrow DB$ for inclusion functor $J:DB_I\hookrightarrow DB$
(this second extension is direct consequence of the first one,
because $J$ does not introduce extension for objects, differently
from $K$, from the fact that $DB_I$ and $DB$ have the same objects).
Thus it is enough to analyze only the first left Kan extension given
by the following commutative diagram:
\begin{diagram}
DB_f      & \rInto^{K} &   DB_I         \\
       & \rdTo^{\Sigma_D}        & \dTo_{Lan_K (\Sigma_D)}\\
       &                &   DB         \\
\end{diagram}
That is, we have the functor $Lan_K:DB^{DB_f} \rightarrow DB^{DB_I}$
is left adjoint to the functor $\_ \circ K:DB^{DB_I}\rightarrow
DB^{DB_f}$, so that left Kan extension of $\Sigma_D \in DB^{DB_f}$
along $K$ is given by functor $Lan_K (\Sigma_D) \in DB^{DB_I}$, and
a natural transformation $\varepsilon:\Sigma_D \rightarrow Lan_K
(\Sigma_D)\circ K$ is an universal arrow. That is, for any other
functor $S:DB_I\rightarrow DB$ and a natural transformation
$\alpha:\Sigma_D \rightarrow S\circ K$, there is a unique natural
transformation $\beta:Lan_K (\Sigma_D) \rightarrow S$ such that
$\alpha = \beta K \bullet \varepsilon$ (where $\bullet$ is a
vertical composition for natural transformations).\\
From the well know theorem for left Kan extension, when we have a
tensorial product $\int^{B \in DB_f}DB_I(K(B),A)\otimes \Sigma_D B$
for every $A \in DB_I$, i.e., $A \in DB$, then the function for
objects of the functor $Lan_K (\Sigma_D)$ is defined by (here $B
\preceq_\omega A$ means that $B \in DB_f ~\& ~B \preceq
A$): \\
$ Lan_K (\Sigma_D)(A) =_{def}\int^{B \in DB_f}DB_I(K(B),A)\otimes
\Sigma_D B\\
= \int^{B \in DB_f}DB_I(B,A)\otimes \Sigma_D B~~~~~~~~~~~~$ (from
$K(D) =D$)\\
$= \int^{B \preceq_\omega A}DB_I(B,A)\otimes \Sigma_D B~ + ~\int^{B
\in DB_f ~\& ~B \npreceq_\omega
A}DB_I(B,A)\otimes \Sigma_D B$\\
$= \int^{B \preceq_\omega A}DB_I(B,A)\otimes \Sigma_D B~ +
~\bot^0~~~~~~~~~~~~$ (hom object $DB_I(B,A)$ is an empty database
$\bot^0$ (zero object in $DB$)  if there is no (monic) arrow from $B$ to $A$) \\
$\simeq \int^{B \preceq_\omega A}DB_I(B,A)\otimes
\Sigma_D B~~~~~~~~~~~~$ (*)\\
$= \int^{B \preceq_\omega A}\widetilde{in_B} \otimes \Sigma_D B~$,
$~~~~~~~$ where $in_B:B \hookrightarrow A $ is unique monic
arrow into $A$\\
$= \int^{B \preceq_\omega A}TB \otimes \Sigma_D
B~$\\
$= \int^{B \preceq_\omega A}TB \otimes T
B~~~~~~~~~~~~$ (for finite $B$, $\Sigma_D(B) = T(B)$)\\
$= \int^{B \preceq_\omega A}TB = \coprod_{B \preceq_\omega A} TB$\\
$ = \bigvee \{T(B) ~|~B \preceq_\omega A \}~~~~~~~~~~$ (lub of compact elements of directed set $\{B ~|~B \preceq_\omega A \}$)\\
$ = T(A)~~~~~~~~~~~~$ (from the fact that the poset $DB_I$ is a
complete algebraic lattice \cite{Majk09a} $(DB_I, \preceq)$ with
meet and join operators $\otimes$ and $\oplus$ respectively, and
with compact elements $TB$ for each \emph{finite} database $B$).
\\
Consequently, we obtain that $ Lan_K (\Sigma_D)$, the extension of
$\Sigma_D$ to all (also infinite non-closed) objects in $DB$, is
equal (up to isomorphism) to endofunctor $T$. That is, formally we
obtain:
   \begin{coro} \label{coro:properties}The following strong connection between
   the relational-algebra signature endofunctor $\Sigma_D$ translated into the database category $DB$ and the closure endofunctor $T$
  hold:  $~~~~ T = \Sigma_D $.
\end{coro}
\textbf{Remark:} Let us consider now which kind of interpretation
can be given to the tensor product (see (*) above): \\$\int^{B \in
DB_f}DB_I(B,A)\otimes \Sigma_D B \simeq \int^{B \preceq_\omega
A}DB_I(B,A)\otimes \Sigma_D B$ and its $B$-components (for $B
\preceq_\omega A$,that is, $B \subseteq_\omega TB \subseteq TA$),
$~~DB_I(B,A)\otimes \Sigma_D B$ in the enriched lfp
database category $DB$:\\
The second component $\Sigma_D B$ can not be set of signature
operators, just because an object in $DB$ can not be set of
functions, and it is not interesting in this interpretation: in fact
it can be omited from $B$-component, because it is equal to $TB$
which is the lub of the first component, i.e., hom object
$DB_I(B,A)= DB_I(B,TA) = \widetilde{in_B}$, for inclusion arrow
$in_B:B \hookrightarrow TA$. \\But for the case when $B
\subseteq_\omega TA$ we have for $f = in_B \bigcup f_B$, where\\
$f_B = \{f_\sigma:B\rightarrow TA ~|~\partial_0(f_\sigma) = B$ and
$\partial_0(f_\sigma) = \{\sigma(\overline{B}) \}$ for each
permutation $\overline{B}$ of relations in $B$ and each operation $\sigma \in \Sigma_R$ with $ar(\sigma) = |B|  \}$, with $\widetilde{f_B} \subseteq TB$,\\
that $DB_I(B,TA) = \widetilde{f} = T(\widetilde{in_B} \bigcup \widetilde{f_B}) = \widetilde{in_B} \oplus \widetilde{f_B} = TB \oplus \widetilde{f_B}$.\\
We can enlarge the source object for $f_B$ to object $TA$ (because
$B \subseteq_\omega TA$), in order to obtain an equivalent mapping
$f_B:TA \rightarrow TA$ and to obtain a representaion of tensor
products by signature operations and signature view-based mappings
$f_B:TA \rightarrow TA$, differently from mappings (i.e. functions)
$\sigma:TA^{ar(\sigma)} \rightarrow TA$ in the standard case when we
use $Set$ category, as the following interpretation:\\$ \int^{B \in
DB_f}DB_I(B,A)\otimes\Sigma_D B\\
 \simeq \int^{B
\subseteq_\omega TA}TB \oplus \widetilde{f_B}$\\
$= \widetilde{f_{\Sigma_D}}$ $~~~~~~~~~~~~$ (where $f_{\Sigma_D}=
(\bigcup_{B
\subseteq_\omega TA} f_B):TA \rightarrow TA$)\\
$= TA~~~~~~~~~~~~~~$ (i.e., $~\Sigma_D(A)~~$ for $~$ \emph{any},
also infinite, database $A \in DB$).\\$\square$\\
%
 As we can see the translation of the relational-algebra
signature
   $\Sigma_R$ is given by the power-view endofunctor $\Sigma_D = T:DB\rightarrow DB$, as
   informally presented in introduction.\\
   Consequently, the endofunctor $(A + \Sigma_D):DB\rightarrow DB$,
from Proposition \ref{prop:cocontinuity}, is the
$\omega$-cocontinuous endofunctor $(A + \Sigma_D):DB\rightarrow DB$,
with a chain\\
$  \perp^0 ~\preceq_0 ~((A + \Sigma_D) \perp^0) ~\preceq_1~ ((A +
\Sigma_D)^2_A \perp^0)~ \preceq_2~~~
 ... ~~~~~~(A + \Sigma_D)^{\omega}  $,\\
 where $(A + \Sigma_D) \perp^0 = \perp^0 + A$, $(A +
\Sigma_D)^2_A \perp^0 = (A + \Sigma_D)(\perp^0 + A) = \perp^0 + A
+TA$, and we obtain that
  the colimit of this diagram in $DB$ is $(A +
 \Sigma_D)^{\omega} = \perp^0 + A + \coprod_\omega TA$. From the fact that for coproduct  (and initial object $\perp^0$)
 holds that $\perp^0 + B \simeq B$ for any $B$, then we can take as  the colimit $(A +
 \Sigma_D)^{\omega} =  A + \coprod_\omega TA$. This colimit is a least fixpoint of the monotone operator $(A + \Sigma_D)$ in a complete lattice
 of databases in $DB$ (Knaster-Tarski theorem).\\
 Notice that the coproduct of two databases $A$ and $B$ in $DB$
 category  \cite{Majk04AOT,Majk08db} corresponds to completely disjoint databases, in the way
 that it is not possible to use relations from these two databases
 in the \emph{same} query: because of that we have that $T(A + B) =
 TA + TB$, that is the set of all views of a coproduct $A + B$ is a
 disjoint union of views of $A$ and views of $B$.\\
  In fact we have that $(A + \Sigma_D)((A +
 \Sigma_D)^{\omega}) = A + T(A + \coprod_\omega TA)  = A +
 TA + T\coprod_\omega TA = A +
 TA + \coprod_\omega TTA = A +
 TA + \coprod_\omega  TA = A + \coprod_\omega TA =
(A + \Sigma_D)^{\omega}$. \\
We can denote this identity arrow in $DB$ category, which is the
initial $(A + \Sigma_D)$-algebra,  by $~~~~[inl_A, inr_A]:(A +
\Sigma_D)(A + \coprod_\omega TA)\rightarrow (A +
\coprod_\omega TA)$.\\
Consequently, the variable injection $inl_A:A \hookrightarrow \T A$
  in $Set$ is translated into  a monomorphism $inl_A:A \hookrightarrow
  (A + \coprod_\omega TA)$ in $DB$ category, with information flux $~\widetilde{inl_A} = TA$.
 The right inclusion
$inr_A:\Sigma_R \T A \hookrightarrow \T A$
  in $Set$ is translated into an isomorphism (which is a monomorphism also)
$~inr_A:\Sigma_D(A + \coprod_\omega TA) \simeq (A + \coprod_\omega
TA)$ in $DB$ category, based on the fact that $\Sigma_D(A +
\coprod_\omega TA) = T (A + \coprod_\omega TA) = TA +
T\coprod_\omega TA = TA + \coprod_\omega TA \simeq A +
\coprod_\omega TA$.\\
So that $~\widetilde{inr_A} = TA + \coprod_\omega TA$ with $TA
\subseteq \widetilde{inr_A}$.\\
   Moreover, by this translation, any $\Sigma_R$ algebra $h:\Sigma_R
   B \rightarrow B$ in $Set$ is translated into an isomorphism
   $h_D:\Sigma_D B \rightarrow B$ with $\widetilde{h_D} = TB$.\\
 Consequently, the initial algebra
 for a given database $A$, with a set of view in $TA$, of the $\omega$-cocontinuous endofunctor $ (A
 + \Sigma_D):DB\longrightarrow DB$ comes with an induction principle,
which we can rephrase the principle as follows: For every
$\Sigma_D$-algebra structure $h_D:\Sigma_DB\longrightarrow B$ (which
must be an isomorphism) and every mapping $f:A\longrightarrow B$
there exists a unique arrow $f_\#:TA\longrightarrow B$ such that the
following diagram in $DB$
\begin{diagram}
A      & \rInto^{inl_A} &   A + \coprod_\omega TA       & \lInto^{~~~~inr_A~~~~} & \Sigma_D (A + \coprod_\omega TA)  \\
       & \rdTo^f        & \dTo_{f_\#}&                & \dTo_{\Sigma_Df_\#}\\
       &                &   B        & \lTo^{h_D}         & \Sigma_D B
\end{diagram}
commutes, where $\Sigma_D = T$ and $f_\# = [f , h_D \circ
\Sigma_Df_\#]$ is the unique \emph{\textbf{inductive extension }of}
$h_D$ \emph{along the mapping} $f$.\\
It is easy to verify that it holds. From the fact that $f = f_\#
\circ inl_A$ we have that $\widetilde{f} = \widetilde{f_\#} \bigcap
\widetilde{inl_A} = \widetilde{f_\#} \bigcap TA =
\widetilde{f_\#}~~\subseteq TA \subseteq \widetilde{inr_A}$. So
there is the unique arrow $f_\#$ that satisfies this condition for a
given arrow $f$. From the fact that $\Sigma_D = T$, we obtain that
$\widetilde{\Sigma_Df_\#} = \widetilde{Tf_\#} = T\widetilde{f_\#} =
\widetilde{f_\#}$ because the information fluxes are closed object
w.r.t. power-view operator $T$. Consequently, we have that
$\widetilde{h_D} \bigcap \widetilde{\Sigma_Df_\#} = TB \bigcap
\widetilde{\Sigma_Df_\#} = \widetilde{\Sigma_Df_\#} =
\widetilde{f_\#} = \widetilde{f_\#} \bigcap TA = \widetilde{f_\#}
\bigcap \widetilde{inr_A}$, so that holds the commutativity $h_D
\circ \Sigma_Df_\# = f_\# \circ inr_A$.\\
The diagram above can be equivalently represented by the following
unique morphism between initial $(A +\Sigma_D)$-algebra and any
other $(A +\Sigma_D)$-algebra:
\begin{diagram}
    A + \coprod_\omega TA       & \lTo^{~~~~[inl_A, inr_A]~~~~} & (A +\Sigma_D) (A + \coprod_\omega TA)  \\
   \dTo_{f_\#}&                & \dTo_{(A + \Sigma_D)f_\#}\\
     B        & \lTo^{[f,h_D]}         & (A +\Sigma_D) B \\
\end{diagram}
Thus we obtain the following Corollary:
\begin{coro} \label{coro:initialsemantics} For each object $A$ in $DB$ category there is the initial
$\Sigma_{A}$-algebra,\\ $<A + \coprod_\omega TA ,~ [inl_A, inr_A]>$,
where $~~inl_A:A \hookrightarrow (A + \coprod_\omega TA)$ is a
monomorphism, while $~~inr_A:\Sigma_D (A + \coprod_\omega TA)
\hookrightarrow (A + \coprod_\omega TA)$ is an isomorphism.
\end{coro}
 This inductive principle can be used to show that the closure
 operator $T$ inductively extends to the endofunctor $T:DB\longrightarrow
 DB$. Indeed, to define its action $Tf$ on arrow $f:A \longrightarrow B$,
 take the inductive extension of $inr_B:\Sigma_D(B + \coprod_\omega TB)\longrightarrow
 (B + \coprod_\omega TB)$ (of the $(B +\Sigma_{D}):DB\longrightarrow DB$ endofunctor with
 initial $(B + \Sigma_R)$-algebra structure $[inl_B , inr_B ]:(B +
 \Sigma_D)(B + \coprod_\omega TB) \longrightarrow (B + \coprod_\omega TB)$ ) along the composite $inl_B \circ
 f$, i.e., $(inl_B \circ
 f )_\# = [inl_B \circ
 f , inr_B \circ \Sigma_D(f+\coprod_\omega Tf)]$.
The second diagram
\begin{diagram}
A      & \rInto^{~~~inl_A~~~} &   A + \coprod_\omega TA        & \lInto^{~~~~inr_A~~~~} & \Sigma_D (A + \coprod_\omega TA)  \\
\dTo^f &                & \dTo_{f+\coprod_\omega Tf}  &                & \dTo_{\Sigma_D(f+\coprod_\omega Tf)}\\
B      & \rInto^{inl_B} &   B + \coprod_\omega TB      & \lInto^{inr_B} & \Sigma_D (B + \coprod_\omega TB)  \\
\end{diagram}
  commutes, thus $(f+\coprod_\omega Tf)\circ inl_A = inl_B \circ f$, so,
  $\widetilde{f+\coprod_\omega Tf}\bigcap \widetilde{inl_A} = \widetilde{Tf}\bigcap
  TA = \widetilde{Tf}= \widetilde{inl_B}\bigcap \widetilde{f}= TB \bigcap
  \widetilde{f} =  \widetilde{f}$, i.e., $\widetilde{Tf} =
  \widetilde{f}$ as originally defined for the endofunctor $T$ (\cite{Majk04AOT}). That $T$ is an endofunctor is easy to verify from the left  commutative
  diagram where the objects $A + \coprod_\omega TA$ can be represented as results of the composed endofunctor
  $E = (I_{DB} + \coprod_\omega \circ T):DB \rightarrow DB$,
  where $I_{DB}$ is the identity endofunctor for $DB$, while  the endofunctor $\coprod_\omega:DB \rightarrow DB$ is a $\omega$ coproduct. \\
  It is easy to verify that $inl_A = \eta_A:A \hookrightarrow EA$, where $EA = A + \coprod_\omega TA$, is obtained from the natural
  transformation $\eta:I_{DB}\longrightarrow E$. Another example
  is the definition of the operation $\mu_A:E^2A \longrightarrow
  EA$ inductively extending $inr_A:\Sigma_D EA \longrightarrow EA$
  along the identity $id_{EA}$ of the object $EA$ (consider the first diagram,
  substituting $A$ and $B$ with the object  $EA = A + \coprod_\omega TA$, $f$ with
  $id_{EA}$ and $h_D$ with $inr_A$). Inductively derived $\eta_A$ (which is a monomorphism),
  $\mu_A$ (which is an identity, i.e., $\mu_A = id_{EA}$, because we have that $E^2 = E$) and the endofunctor $E$, define the monad $(E, \eta ,
  \mu )$, i.e., this monad is inductively extended in a \emph{natural
  way} from the signature endofunctor $\Sigma_{D}= T:DB\longrightarrow
  DB$.\\
  Thus, the monad $(E, \eta ,  \mu)$, where $E = I_{DB} + \coprod_\omega \circ T = I_{DB} + T \circ \coprod_\omega$  is an inductive algebraic extension
  of the "coalgebraic" \emph{observation based} power-view monad $(T, \eta, \mu)$.
\subsection{A coalgebraic view: corecursion and infinite trees}
Coalgebras are suitable mathematical formalizations of reactive
systems and their behavior, like to our case when we are considering
databases from the query-answering, that is, view-based approach.\\
This subsection presents an application of corecursion, that is, of
construction method using final coalgebras ~\cite{AAMV01}. In order
to better understand the rest lets give an example for a
coalgebraic point of view of a database mappings. \\
\textbf{Example:} Let a database $A$ contain two relations, $r_P$
(of a predicate $P$ with 4 attributes), and $r_Q$ (of the predicate
$Q$ with 5 attributes), such that 4-th attribute of $r_P$ and 3-th
attribute of $r_Q$ are of the same domain. Let define the mapping at
logical level from $A$ to $B$, which contains the relation $r_R$ (of
a predicate $R$ with two attributes) by the conjunctive query
$R(x,y) \leftarrow P(a,x,z)\wedge Q(b,y,z)$, (which by completion is
an equivalence, $R(x,y) \leftrightarrow P(a,x,z)\wedge Q(b,y,z)$)
where 'a','b', are constants of a domain and $x,y,z$ are attribute
variables.\\
Let us define now a relational algebra signature, $\Sigma$, with
sorts correspondent to tuples of variables which represent the
views and (also infinite) set of unary and binary basic operations:\\
$\Sigma = Op_1 + Op_2$, where\\
$Op_1 = \{ \pi_{(S)}~|~S \in \P(N)$ for some finite $N \}
~\bigcup~\{ Where_{(C)}~|~$ C is any selection condition on
attributes $\}$\\
$Op_2 = \{ Join_{(v = w)}~|~$ v, w are relation's attributes $ \}
~\bigcup~\{ Union \}$\\
Now, the mapping $f:A\rightarrow B$ given by the logic implication
above my be equivalently expressed by the following system of
guarded equations:\\
$~~~~~~~~~~<x,y> \approx \pi_{(2,5)}(<v_1,x,z,w_1,y,z_1>)$\\
 $~~~~~~~~~~<v_1,x,z,w_1,y,z_1> \approx Join_{(v =
w)}(<v_1,x,z>,<w_1,y,z_1>)$\\
$~~~~~~~~~~<v_1,x,z> \approx Where_{(v = 'a')}(<v,x,z>)$\\
$~~~~~~~~~~<w_1,x,z_1> \approx Where_{(w = 'b')}(<w,x,z_1>)$\\
$~~~~~~~~~~<v,x,z> \approx \pi_{(2,3,4)}(<x_1,v,x,z>)$\\
$~~~~~~~~~~<w,x,z_1> \approx \pi_{(1,2,3)}(<w,x,z_1,z_2,z_3>)$\\
$~~~~~~~~~~<x_1,v,x,z> \approx r_P$\\
$~~~~~~~~~~<w,x,z_1,z_2,z_3> \approx r_Q$\\
such that the relation $r_R$ is the solution of this system for the
tuple-variable $<x,y>$.\\ The polynomial endofunctor of $Set$,
$H_\Sigma:Set\rightarrow Set$, derived by this signature $\Sigma$,
for any given set of tuple variables $X$ (in example above $<x,y>,
<v_1,x,z>, <w_1,x,z_1>, <v,x,z>, <w,x,z_1>, <x_1,v,x,z>,
<w,x,z_1,z_2,z_3>, <v_1,x,z,w_1,y,z_1>
\in X$), is of the form \\
$H_\Sigma(X) = \coprod_{n< \omega} Op_n\times X^n = \coprod_{n \in
\{1,,2\}} Op_n\times X^n$\\
It is easy to verify that right parts of equations (except two last
equations) belong to $H_\Sigma(X)$. The right parts of the last two
equations belong to "parameters" database $A$, i.e., $r_P,r_Q
\in A$.\\
Thus, the system of guarded equations above, which define a mapping
from a database $A$ to a database $B$, my be expressed by the
function $f_e:X\rightarrow H_\Sigma(X)+A$ (for example,
$f_e(<v_1,x,z,w_1,y,z_1>) = Join_{(v = w)}(<v_1,x,z>,
<w_1,y,z_1>)$), which is just a coalgebra of the polynomial $Set$
endofunctor $H_\Sigma(\_)+A:Set\rightarrow Set$ with the signature
$\Sigma_X = \Sigma \bigcup X$ (the tuple-variables
in $X$ are seen as operations of arity 0). \\
It is known ~\cite{AAMV01} that such polynomial endofunctors of
$Set$ have a \textbf{final coalgebra} which is the algebra of all
finite and infinite $\Sigma_X$-labelled trees, i.e., the set of all
views $T_\infty(A)$,  so that  $T_\infty(A) =
H_\Sigma(T_\infty(A))+A~~$, i.e., $T_\infty(A)$ is the maximal
fixpoint of the endofunctor
$H_\Sigma(\_)+A$.\\
So we obtained that for any database $A$, its complete power-view
object $T_\infty(A)$ corresponds to the final coalgebra of the
iteratable endofunctor $H_\Sigma(\_)+A$, in the way that the guarded
system of equations defined by a database mapping $f$ has the
\textbf{unique} solution $s:X\rightarrow T_\infty(A)$, which is the
$H_\Sigma(\_)+A$-coalgebra homomorphism from the coalgebra $(X,f_e)$
into the final coalgebra $(T_\infty(A),\simeq)$, as given by the
following commutative diagram in $Set$:
\begin{diagram}
X      & \rTo^{s} &   T_\infty(A ) \\
\dTo^{f_e} &                & \dTo_{\simeq}  \\
H_\Sigma(X)+A      & \rTo^{H_\Sigma(s)+A} &   H_\Sigma(T_\infty(A))+A  \\
\end{diagram}
It means, for example, that $s(<x,y>) \in T_\infty(A)$ is the unique
solution of the conjunctive formula $P(a,x,z)\wedge Q(b,y,z)$ which
is given in the body of the mapping query from a database $A$ into a
database $B$, and which is a part of the minimal Herbrand model for
the logic theory expressed by this database mapping.\\ Let us now
consider coalgebra properties in $DB$ category.We define an
\emph{iteratable} endofunctor $H$ of a category $\D$ if for every
object $X$ of $\D$ the endofunctor $H(\_) + X$ has an final algebra.
We are going to show that the signature endofunctor  $\Sigma_R$ is
iteratable.
\begin{propo} \label{def:finalsemantics} Every endofunctor $\Sigma_{R_A}= (\Sigma_R + A):DB\longrightarrow
DB$ has the final $\Sigma_{R_A}$-coalgebra, $<T_\infty A,~ <p_l ,
p_r >:T_\infty A \rightarrow \Sigma_RT_\infty A + A>$, where
$p_l:T_\infty A \twoheadrightarrow A$ and $p_r:T_\infty A
\twoheadrightarrow \Sigma_R T_\infty A $ are the unique product
epimorphisms of the (co)product $TA \simeq \Sigma_RTA + A$ obtained
as a maximal fixpoint of this endofunctor. Thus for any database $A$
its power-view object $T_\infty A$, that is the set of \textbf{all}
views of $A$ obtained by \textbf{finite and infinite} tree terms of
the $SPJRU$ relational algebra, is a final $\Sigma_{R_A}$-algebra.
\end{propo}
The final coalgebra $<T_\infty A, <p_l , p_r >>$ (where $<p_l , p_r
>$ is an isomorphism) of the endofunctor $\Sigma_{R_A}= \Sigma_R +
A:DB\longrightarrow DB$ comes with an coinduction principle, and
since it is the (co)product $ \Sigma_RTA + A$, we can rephrase the
principle as follows: For every $\Sigma_R$-coalgebra structure
$h:B\rightarrow \Sigma_RB$ (which is an isomorphism) and every
mapping $f:B\longrightarrow A$ there exists a unique arrow
$f^\#:B\rightarrow T_\infty A$ such that the diagram
\begin{diagram}
A      & \lOnto^{p_l} &   T_\infty A       & \rOnto^{p_r} & \Sigma_R T_\infty A  \\
       & \luTo^f        & \uTo_{f^\#}&                & \uTo_{\Sigma_Rf^\#}\\
       &                &   B        & \rTo^h         & \Sigma_R B \\
\end{diagram}
commutes in $DB$, where $f^\# = <f, \Sigma_Rf^\# \circ h>$ is the
unique \emph{\textbf{coinductive extension }of} h \emph{along the
mapping}f.\\
Note that $f^\#:<B, <h , f>> \rightarrow <T_\infty A, <p_l , p_r
>>$ is the unique arrow to the final
$\Sigma_{R_A}$-coalgebra from the coalgebra of the map $<h ,
f>:B\rightarrow \Sigma_R B + A$:
\begin{diagram}
B      & \rTo^{f^\#} &   T_\infty(A ) \\
\dTo^{<h,f>} &                & \dTo_{\simeq}  \\
\Sigma_R(B)+A      & \rTo^{\Sigma_R(f^\#)+A} &   \Sigma_R(T_\infty(A))+A  \\
\end{diagram}
 This coinductive principle can be used to show that the closure
 operator $T_\infty$ coinductively extends to the endofunctor $T_\infty :DB\longrightarrow
 DB$. Indeed, to define its action $T_\infty f$ on arrow $f:B \longrightarrow A$,
 take the inductive extension of $p_r:T_\infty B \rightarrow \Sigma_RT_\infty B$
 (of the $\Sigma_{R_B}:DB\longrightarrow DB$ endofunctor with
 the final $\Sigma_{R_B}= (\Sigma_R + B)$-coalgebra structure $<p_l , p_r >:T_\infty B \rightarrow
 \Sigma_RT_\infty B + B$ ) along the composite $f \circ
 p_l$, i.e., $T_\infty f \triangleq (f \circ
 p_l )^\# = <f \circ p_l , \Sigma_RT_\infty f \circ p_r>$. (Note that $T_\infty f$ can be seen as a
 homomorphism from the $\Sigma_R$-coalgebra $<T_\infty B, p_r>$ to the $\Sigma_R$-coalgebra $<T_\infty A,
 p_r>$).\\
So we obtain the following commutative diagram in $DB$:
\begin{diagram}
A      & \lOnto^{p_l} &   T_\infty A       & \rOnto^{p_r} & \Sigma_R T_\infty A  \\
\uTo^f &                & \uTo_{T_\infty f}  &                & \uTo_{\Sigma_RT_\infty f}\\
B      & \lOnto^{p_l} &   T_\infty B       & \rOnto^{p_r} & \Sigma_R T_\infty B  \\
\end{diagram}
  Thus, final coalgebras of the functors
  $\Sigma_{R_A}$ form a monad $(T_\infty, \eta ,  \mu)$, called the
  \emph{completely iterative monad} generated by signature $\Sigma_R$.
\section{Conclusions}
In previous work we defined a category $DB$ where objects are
databases and morphisms between them are extensional GLAV mappings
between databases. We defined equivalent (categorically isomorphic)
objects (database instances) from the \emph{behavioral point of view
based on observations}:  each arrow (morphism) is composed by a
number of "queries" (view-maps), and each query may be seen as an
\emph{observation} over some database instance (object of $DB$).
Thus, we  characterized each object in $DB$ (a database instance) by
its behavior according to a given set of observations. In this way
two databases $A$ and $B$ are equivalent (bisimilar) if they have
the same set of its observable internal states, i.e. when $TA$ is
equal to $TB$. It has been shown that such a $DB$ category is equal
to its dual, it is symmetric in the way that the semantics of each
morphism is an closed object (database) and viceversa each database
can be represented by its identity morphism, so that $DB$ is a
2-category.\\
 In \cite{Majk09a,Majk09a} has been introduced the categorial (functors)
semantics for two basic database operations: \emph{matching} and
\emph{merging} (and data federation), and has been defined the
algebraic database lattice. In the same paper has bee shown that
$DB$ is concrete, small and locally finitely presentable (lfp)
category, and that $DB$ is also monoidal symmetric V-category
enriched over itself. Based on these results  the authors developed
a metric space and a subobject classifier for $DB$ category, and
they have shown that it is a weak monoidal topos.\\
In this paper we presented some other contributions for this
intensive exploration of properties and semantics of $DB$ category.
Here we considered some Universal algebra considerations and
relationships of $DB$ category and standard $Set$ category.
  We defined  a categorial coalgebraic semantics for GLAV database mappings
  based on monads, and of general (co)algebraic and (co)induction
  properties for databases. \\
  It was shown that a categorial semantics of database mappings can
  be given by the Kleisly category of the power-view monad $T$, that
  is, was show that Kleisly category is a model for database
  mappings up to the equivalence $\approx$ of morphisms in $DB$
  category. It was demonstrated that Kleisly category is isomorphic
  to the $DB$ category, and that call-by-values and call-by-name
  paradigms of programs (database mappings) are represented by
  equivalent morphisms. Moreover, it was shown that each database
  query (which is a program) is a monadic $T$-coalgebra, and that
  any morphism between two $T$-coalgebras defines the semantics for
  the relevant query-rewriting.


\bibliographystyle{IEEEbib}

\bibliography{medium-string,krdb,mydb}


\end{document}